\documentclass[10pt]{article}

\usepackage{amsmath}

\usepackage{array}

\usepackage{appendix}

\usepackage{tocloft}                   

\usepackage{graphicx}

\usepackage{amsfonts}

\usepackage{amssymb}

\usepackage{mathrsfs}

\usepackage{yfonts}

\usepackage{euscript}

\usepackage{centernot}                 

\usepackage{ifsym}                     

\usepackage{upgreek}

\usepackage{mathtools}

\usepackage{color}

\usepackage{slantsc}
\usepackage{calligra}

\usepackage{bbold}          

\usepackage[T1]{fontenc}

\usepackage{epsf}

\usepackage{latexsym}

\usepackage{tipa}

\usepackage{makeidx}

\makeindex

\def\b{\begin{equation}}

\def\e{\begin{equation}}

\def\be{\begin{equation}}              

\def\ee{\end{equation}}

\def\beq{\begin{equation}}

\def\eeq{\end{equation}}

\def\bea{\begin{eqnarray}}

\def\eea{\end{eqnarray}}

\def\m{\mbox{ }}

\def\mma {\m , \m \m }

\def\!{\hspace{-1.6667em}}

\def\Proof{{\n{\u{Proof}}}\m}

\def\c{\cite}

\def\l{\label}

\def\r{\ref}

\def\n{\noindent}

\def\f{\footnote}

\def\u{\underline}

\def\s{\stackrel}

\def\mB{\mbox{B}}  

\def\mC{\mbox{C}}                        

\def\mD{\mbox{D}}                        

\def\mF{\mbox{F}}

\def\mG{\mbox{G}}

\def\mK{\mbox{K}}

\def\mO{\mbox{O}}

\def\mP{\mbox{P}}

\def\mS{\mbox{S}}                        

\def\mT{\mbox{T}} 

\def\mV{\mbox{V}}

\def\mW{\mbox{W}}

\def\me{\mbox{e}}

\def\ml{\mbox{l}}

\def\mo{\mbox{o}}

\def\mp{\mbox{p}}

\def\se{\mbox{\scriptsize e}}

\def\sf{\mbox{\scriptsize f}}

\def\sm{\mbox{\scriptsize m}}

\def\sr{\mbox{\scriptsize r}}

\def\sG{\mbox{\scriptsize G}}

\def\sO{\mbox{\scriptsize O}}

\def\sT{\mbox{\scriptsize T}}

\def\sV{\mbox{\scriptsize V}}

\def\sfX{\mbox{\sffamily{\scriptsize X}}}      

\def\sumi2{\sum\mbox{}_{\mbox{}_{\mbox{\scriptsize $i$=1}}}^2}

\def\sumi3{\sum\mbox{}_{\mbox{}_{\mbox{\scriptsize $i$=1}}}^3}

\def\sumABcycles3{\sum\mbox{}_{\mbox{}_{\mbox{\scriptsize cycles $A,B$=1}}}^{3}}

\def\sumCDcycles3{\sum\mbox{}_{\mbox{}_{\mbox{\scriptsize cycles $C,D$=1}}}^{3}}

\def\sumj3{\sum\mbox{}_{\mbox{}_{\mbox{\scriptsize $j$=1}}}^3}

\def\sumk3{\sum\mbox{}_{\mbox{}_{\mbox{\scriptsize $k$=1}}}^3}

\def\prodiA1{\prod\mbox{}_{\mbox{}_{\mbox{\scriptsize $i$=1}}}^{A - 1}}

\def\bigtimes{\mbox{\Large $\times$}}

\def\es{\m = \m}

\def\:={\m := \m}

\def\=:{\m =: \m}

\def\geqs{\m \geq \m}

\def\FrI{\mbox{$\mathfrak{I}$}}                                

\def\FrA{\mbox{$\mathfrak{A}$}}                                

\def\FrE{\mbox{$\mathfrak{E}$}}                                

\def\FrT{\mathfrak{T}}                                         

\def\FrC{\mbox{$\mathfrak{C}$}}                                
                                                       
\def\sFrR{\mbox{\scriptsize $\mathfrak{R}$}}                   
\def\FrE{\mbox{$\mathfrak{E}$}}                                

\def\FrS{\mbox{\Large $\mathfrak{s}$}}                         
\def\sFrS{\mbox{\large$\mathfrak{s}$}}                         

\def\lFrg{\mbox{\Large$\mathfrak{g}$}}                         
\def\FrT{\mbox{\boldmath$\mathfrak{T}$}}                       
\def\Hilb{\mbox{{\boldmath$\mathfrak{H}$}ilb}}                 
\def\FrQ{\mbox{\Large $\mathfrak{q}$}}                               
\def\bFrL{\mbox{\boldmath$\mathfrak{L}$}}                            

\def\Phase{\mbox{{\boldmath$\mathfrak{P}$}hase}}                     

\def\bFrR{\mbox{\boldmath$\mathfrak{R}$}}                            
\def\Rig-Phase{\bFrR\mbox{ig-}\Phase}                                
                                                                													   
\def\FrP{\mbox{\Large $\mathfrak{p}$}}                                 
\def\FrR{\mbox{\boldmath$\mathfrak{R}$}}                             

\def\sFrR{\mbox{\scriptsize\boldmath$\mathfrak{R}$}}                 
\def\1mat{\u{\u{1}}}                                                 

\def\Positive-Modespace{\mbox{{\boldmath$\mathfrak{M}$}odespace$^+$}}

\def\POSITIVE-MODESPACE{\mbox{{\boldmath$\mathfrak{M}$}ODESPACE$^+$}}

\def\Leib{\bFrL\mbox{eib}}                                           

\def\Top{\FrT\mo\mp}

\def\Rel{\FrR\me\ml}

\def\Kin-Hilb{\mbox{{\boldmath$\mathfrak{K}$}in-\Hilb}}                     

\def\Mid-Hilb{\mbox{{\boldmath$\mathfrak{M}$}id-\Hilb}}                     

\def\Dyn-Hilb{\mbox{{\boldmath$\mathfrak{D}$}yn-\Hilb}}                     

\def\5Star{\mbox{\Large$\star$}}              

\def\NSI{Na\"{\i}ve Schr\"{o}dinger Interpretation }

\begin{document}

\begin{titlepage}

\begin{center}

\large{\bf TOPOLOGICAL SHAPE THEORY} \normalsize

\vspace{0.1in}

\normalsize

\vspace{0.1in}

{\large \bf Edward Anderson$^*$}

\vspace{.2in}

\end{center}

\begin{abstract}

\n Kendall's Similarity Shape Theory for constellations of points in the carrier space $\mathbb{R}^n$ was developed for use in Probability and Statistics. 
It was subsequently shown to reside within (Classical and Quantum) Mechanics' Shape-and-Scale Theory, 
in which the points are interpreted as particles and the carrier space plays the role of absolute space.  
In other more recent work, Kendall's Similarity Shape Theory has been generalized to affine, projective, conformal and supersymmetric versions, 
as well as to $\mathbb{T}^n$, $\mathbb{S}^n$, $\mathbb{RP}^n$ and Minkowski spacetime carrier spaces. 
This has created a sizeable field of study: generalized Kendall-type Geometrical Shape(-and-Scale) Theory.  
Aside from offering a wider range of shape-statistical applications, this field of study is an exposition of models of Background Independence 
of relevance to the Absolute versus Relational Motion Debate, and the Foundations and Dynamics of General Relativity and Quantum Gravity.

\m 

\n In the current article, we consider a simpler type of Shape Theory, comprising relatively few types of behaviour: the Topological Shape Theory of rubber shapes. 
This underlies the above much greater diversity of more structured Shape Theories; 
in contrast with the latter's (stratified) manifolds shape spaces, the former's are just graphs. 
We give examples of these graphs for the smallest nontrivial point-or-particle numbers, 
and outline how these feature within a wider range of Geometric Shape Theories' shape spaces, 
and are furthermore straightforward to do Statistics, Dynamics and Quantization with. 

\end{abstract}

\n {\bf PACS}: 04.20.Cv, 02.40.Pc. {\bf Physics keywords}: background independence, configuration spaces, dynamical and quantization aspects of General Relativity.

\m 

\n {\bf Mathematics keywords}: shapes, spaces of shapes, Shape Topology, Shape Statistics, Automorphisms, Applied Graph Theory.

\vspace{0.1in}
  
\n $^*$ Dr.E.Anderson.Maths.Physics@protonmail.com

\vspace{0.1in}
 
{            \begin{figure}[!ht]
\centering
\includegraphics[width=1\textwidth]{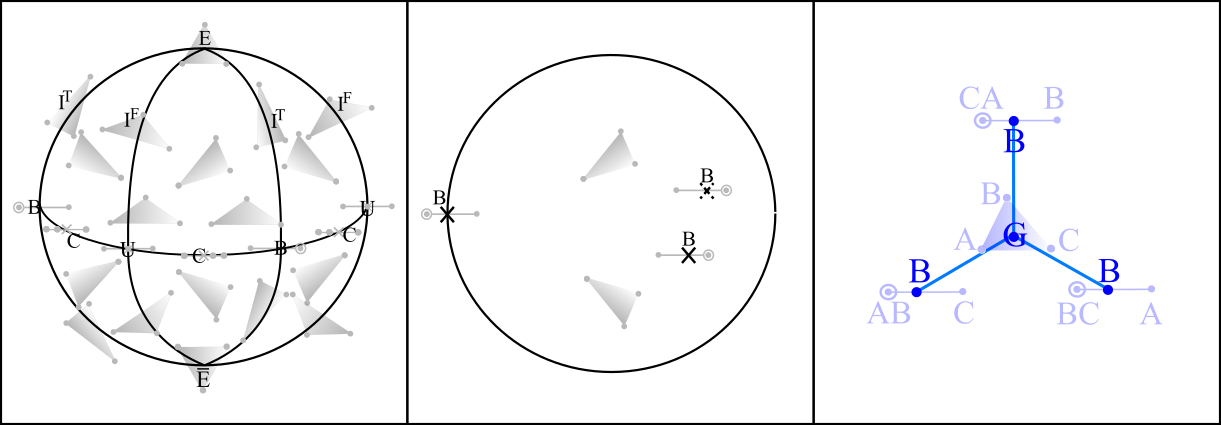}
\caption[Text der im Bilderverzeichnis auftaucht]{        \footnotesize{a) The triangleland -- i.e.\ 2-$d$ 3-point-or-particle shape space -- sphere.  
Its poles E are each of the two labelling orientations of equilateral triangles, 
the meridians marked I are isosceles triangles and the equator C is where collinear configurations are located.
The B points are binary coincidences, for which two of the triangle's three points are superposed. 
The U points are maximally uniform collinear shapes.   
$\sfX$ denotes here the centre of mass of the outermost 2 points-or-particles.  

\m

\n b) The metric-level configuration space's topological features.

\m

\n c) The deep blue drawing here is of the triangleland topological claw graph underlying b) and a) (and other theories of shape(-and-scale) in dimension $\geq 2$.
Pale blue is used for the corresponding rubber shapes, with A, B, C being distinguishable particle labels. 
c)'s greater simplicity relative to a) is part of the reason for the current article.
The other point is that a single graph like c) occurs multiple times throughout the rich complex of models that constitute Shape(-and-Scale) Theory.} }
\l{S(3, 2)-Intro} \end{figure}          }
 
\end{titlepage}

\section{Introduction}\l{Intro-I}

\n Shape Theory in David Kendall's sense \c{Kendall84, Kendall89, Kendall} models space as $\mathbb{R}^d$ of dimension $d$, 
and treats constellations of $N$ points thereupon by quotienting out the similarity group of transformations, $Sim(d)$.   
This involves a metric notion of shape, and concentrates on the configuration space formed by these metric shapes: {\it shape space}, 
\be 
\FrS(N, \, d)   \:=  \frac{\bigtimes_{i = 1}^N \mathbb{R}^d}{Sim(d)}  
             \es  \frac{\mathbb{R}^{N \, d}}{Sim(d)}                  \m .  
\label{Sim-S}			 
\ee 
This work remains rather more familiar in the Shape Statistics literature \c{Small, Kendall, Bhatta, DM16, PE16}, 
in which probability measures and statistics are set up in concordance with the shape space's geometry.  

\m 

\n See moreover e.g.\ \c{LR95, LR97, ArchRat, Montgomery2, AF, +Tri, FileR, QuadI, APoT2, QuadII, AKendall, ABeables, ABeables2, MIT, ABeables3, ABook, 
I, II, III, A-Pillow, 2-Herons, Max-Angle-Flow, Ineq} for related work in other fields, 
including Mechanics, Quantization and modelling some aspects of Classical and Quantum General Relativity's Background Independence \cite{A64, A67, I93, Giu06, RovelliBook}.
Some of these further works consider quotienting instead by the Euclidean group of transformations, $Eucl(d)$. 
This gives Shape-and-Scale Theory. 
The corresponding configuration space for this is {\it relational space} alias {\it shape-and-scale space}, 
\be 
\FrR(N, \, d)  \es  \frac{\mathbb{R}^{N \, d}}{Eucl(d)} 
                 = \mC(\FrS(N, \, d))                                      \m ;  
\label{Eucl-R}				 
\ee 
this is meant in the sense of the Absolute versus Relational Debate \c{Newton, L, LCC, M, BB82, DoD, Buckets, ABook} which dates at least as far back as Newton versus Leibniz.   
In physical applications, the points are often considered to be particles, so we use `point-or-particle' as a portmanteau name and concept.
In the case of statistical applications, the points represent location data.  
The last equality in  (\r{Eucl-R}) indicates that this turns out to be equivalent to taking the cone $\mC$ over the shape space, 
where the extra degree of freedom thus included is the scale variable.  

\m 

\n Three further ambiguities in the above modelling, substantially further enriching both its foundational scope and its applicability, are as follows.  
These ambiguities have created a large and rich field of study: generalized Kendall-type Geometrical Shape(-and-Scale) Theory.  

\m 

\n 1) $\mathbb{R}^d$ plays the role of absolute space, or, more generally of a carrier space (i.e.\ a not necessarily physically realized counterpart). 
This role can be allotted to other models of space, such as $\mathbb{S}^d$, $\mathbb{T}^d$ and $\mathbb{RP}^d$.    

\m 

\n 2) The role of the continuous group being quotiented out is that of a group of automorphisms that are held to be irrelevant to the modelling in question.  
This implements part of what theoretical physicists term Background Independence. 
Alternatives to $Sim(d)$ and $Eucl(d)$ here include  
the affine group \cite{Sparr, GT09, Bhatta, PE16}, 
its scaled counterpart the equi-top-voluminal group \cite{AMech, ABook}, 
or the conformal group \cite{AMech, ABook}.  
With a bit more work, projective \cite{MP03, Bhatta, PE16} 
and supersymmetric \cite{AMech, ABook} options are also available, 
as are all compatible combinations of these transformations.  

\m 

\n 3) One can furthermore quotient out by discrete transformations to model such as mirror image indistnguishability or point-or-particle label indistinguishability 
(including {\sl partial} label indistinguishablilty \cite{I, II, III}).  

\m 

\n A more detailed account of 1), 2) and 3) is given in Sec 2.  
The current Article moreover concentrates rather on a coarser view universal within the above pletora of Shape(-and-Scale) Theories: 
the Topological Shape(-and-Scale) Theory we introduce in Sec 3.  
This maintains significant distinction between the $d = 1$ and $d > 1$ versions, but has no further dependence on spatial dimension.  
We provide $(N, \, d)$ = (3, 1), (4, 1), (3, 2), (4, 2) examples to illustrate Topological Shape(-and-Scale) Theory in Secs 4 to 8 respectively.  
To Geometric Shape(-and-Scale) Theories' reduced configuration spaces being stratified manifolds in general, Topological Shape(-and-Scale) Theory's are just graphs.
We shall see that for $d > 1$, these graphs' order is closely related to the number of partitions of $N$, 
but for $d = 1$ it exceeds that number, reflecting a more fine-grained structure.  

\m

\n N.B.\ that in addition to many of these examples being used in \c{I, II, III, IV} for $\mathbb{R}^d$ carrier space, 
they are now revealed to also underlie 1) and 2)'s variants.  
The full extent of (4, 1) and (4, 2)'s variant 3) is moreover new to the current article, even for the $\mathbb{R}^d$ $Sim(d)$ and $Eucl(d)$ setting.  

\m 

\n As applications of the current article's Topological Shape(-and-Scale) Theory, 
we firstly outline how these feature within a wider range of Geometric Shape(-and-Scale) Theories' shape(-and-scale) spaces in Sec 9.
We next outline that Probability and Statistics are much easier to formulate on graphs than on (stratified) manifolds in Sec 10, 
and consider Dynamics and Quantization on shape(-and-scale) graphs in Sec 11.  
We conclude in Sec 12 with commentaries on both universal usefulness over 1), 2) and 3) and Background Independence modelling.

\vspace{10in}

\section{Geometrical Shape(-and-Scale) Theories}

\n{\bf Definition 1} {\it Carrier space} $\FrC^d$, alias absolute space in the physically realized case,  
is an at-least-provisional model for the structure of space.

\m

\n{\bf Remark 1} See below for four different examples of carrier spaces, and \cite{I, A-Generic, Project-1, A-Local} for further discussion.\footnote{The second of these
articles considers the generic case, the third special-and-general relativistic spacetime analogues and the fourth, the locally-approximate case.}

\m 

\n{\bf Modelling Assumption} The current article restricts itself to carrier spaces which are connected manifolds without boundary.

\m

\n{\bf Definition 2} {\bf Constellation space} is the product space 
\be 
\FrQ(N, \FrC^d) = \bigtimes_{i = 1}^N \FrC^d                \m .  
\ee 

\m 

\n{\bf Definition 3}  A {\it relational theory} is a quadruple 
\be
(\FrC^d, N, \lFrg, \Gamma)
\ee 
for $\FrC^d$ a carrier space, 
$N$ a point-or-particle number, 
$\lFrg$ a continouous group of automorphisms acting on $\FrC^d$, and 
$\Gamma$ a discrete group of automorphisms acting on $\FrQ(N, \FrC^d)$.
$\lFrg$ is in more detail $Aut(\langle\FrQ, \sigma\rangle)$, 
for $\sigma$ some level of mathematical structure on $\FrQ$ which is itself preserved by the automorphisms in hand.\f{This covers both $\lFrg$ and $\Gamma$. 
More generally, one could have an unsplittable group playing a joint role running over both of these.}
%
For now, $\Gamma = id$ so $Aut(\langle \FrQ, \sigma \rangle)$ is just $\lFrg$.
$\FrQ$'s product form allows $\lFrg$ to be recharacterized as $Aut(\langle \FrC^d, \sigma \rangle)$.  

\m

\n{\bf Definition 4} {\it Relationalspace} is the quotient space
\be  
\Rel(\FrC^d, N, Aut(\FrQ, \sigma)) \m = \m \frac{\FrQ(N, \FrC^d)}{Aut(\langle\FrQ, \sigma\rangle)}       \m .  
\ee
\n{\bf Remark 2} Splitting into continuous and discrete automorphisms (where possible), 
\be
\Rel(\FrC^d, N; \lFrg, \Gamma) = \frac{  \Rel(N, \, d)  }{  \lFrg \circ \Gamma  }   \m .
\ee
Here $\circ$ is a generic product (of the form $\times$ or $\rtimes$ -- semidirect product of groups -- in all examples in the current article).

\m 

\n{\bf Remark 3} Within this scheme,  
\be 
\Rel(\FrC^d, N; id , id)   \es  \frac{\FrQ(\FrC^d, N)  }{  id \times id  } 
                           \es  \frac{\FrQ(\FrC^d, N)  }{  id  }
                           \=:  \FrQ(\FrC^d, N)                             \m :  
\ee
constellation space itself.  

\m

\n{\bf Definition 5} For those $\lFrg$ that do not include a scaling transformation, 
the relationalspace notion specializes to the {\it shape space} notion \cite{Kendall84, Kendall, FileR, AMech, PE16, A-Monopoles} 
\be 
\FrS(N, d; \Gamma) := \Rel(d, N; \emptyset, \Gamma)   \m . 
\ee
\n{\bf Definition 6} For those $\lFrg$ that do include a scaling transformation, 
the relationalspace notion specializes to {\it shape-and-scale space} notion \cite{LR95, FileR, AMech, ABook, A-Monopoles} 
\be 
\FrR(N, d; \Gamma) := \Rel(d, N; s \Gamma)  \m .
\ee
\n{\bf Remark 4} Relational Theory is thus a portmanteau of Shape Theory and Shape-and-Scale Theory.  
The distinction of whether or not scaling is among the automorphisms is significant in practise because many of the most-studied models are part of a 
{\it shape space and shape-and-scale-space pair}.
This corresponds to Shape Theories which remain algebraically consistent upon removal of an overall dilation generator. 
However, more generally there are plenty of instances of singletons, as we shall see below.  
The first five examples below are moreover all for flat Euclidean carrier spaces $\mathbb{R}^d$.  

\m 

\n{\bf Example 1} Quotienting out by the similarity group $Sim(d)$ gives {\it Kendall's Similarity Shape Theory} \cite{Kendall84}.
In particular, in 1- and 2-$d$ for $\Gamma = id$, the shape spaces 
\be 
\FrS(N, \, d) := \FrS(N, \mathbb{R}^d)
\ee 
for this are the spheres $\mathbb{S}^{N - 2}$ in 1-$d$ and complex projective spaces $\mathbb{CP}^{N - 2}$ in 2-$d$.  

\m 

\n{\bf Example 2}  Quotienting out by the Euclidean group $Eucl(d)$ gives {\it Metric Shape-and-Scale Theory}.
It is quite likely that Leibniz would have liked to have such a theory available for its implementation of the relational side of the specific setting.
In particular, for $\Gamma = id$, the shape-and-scale spaces for this are
the real spaces $\mathbb{R}^{N - 1}$ in 1-$d$ and cones over complex projective spaces $\mC(\mathbb{CP}^{N - 2})$ in 2-$d$.  

\m 

\n{\bf Remark 5} Examples 1 and 2 moreover constitute a first shape and shape-and-scale pair.  

\m 

\n{\bf Example 3} Quotienting out by the affine group $Aff(d)$ gives {\it Sparr's Affine Shape Theory} \cite{Sparr, GT09, AMech, PE16}.
This requires $d > 1$ to be distinct from the similarity case, and is first nontrivial for $N = 4$.  
The corresponding affine shape spaces are  
\be 
\FrA(N, \, d)  :=   \FrA(N, \mathbb{R}^d)  
            \es  \frac{\mathbb{R}^{N \, d}}{Aff(d)}  \m .  
\ee
These are moreover found to be stratified manifolds \cite{GT09} whose constituent strata are Grassmann spaces.    
A major application of this Affine Shape Theory is Image Analysis; in particular $\FrA(N, \, d)$ is the space of $N$-point images in dimension $d$ as viewed from infinity.  

\m  

\n{\bf Example 4} Quotienting out by the equi-top-voluminal group $Aff(d)$ gives {\it Equi-top-voluminal Shape Theory} \cite{Sparr, GT09, AMech, PE16}.
This is first distinct for $d = 2$, for which the most familiar example of the underlying spatial geometry -- equiareal geometry \cite{Coxeter} -- occurs; 
again, this is first nontrivial for $N = 4$.    
The corresponding equi-top-voluminal shape-and-scale space is 
\be 
\FrE(N, \, d)  \es  \frac{\mathbb{R}^{N \, d}}{Equi(d)}  \m .  
\ee 
This is also a stratified manifold whose strata are now cones over Grassmann spaces \cite{Affine-Shape-1, Affine-Shape-2}.    

\m 

\n{\bf Remark 6} Examples 3 and 4 constitute a second shape and shape-and-scale pair.  

\m 

\n{\bf Example 5} Quotienting out by the conformal group $Conf(d)$ gives {\it Conformal Shape Theory} 
The corresponding conformal shape spaces are
\be 
\FrC(N, \, d)  \es  \frac{\mathbb{R}^{N \, d}}{Conf(d)}  \m .  
\ee
In the sense paralleling Kendall's, \cite{AMech} remains largely unexplored due to being less technically straightforward to handle.  

\mbox{ }

\n{\bf Complication 1} Conformal Shape Theory is only meaningfully defined in dimension $\geq 3$. 
[For $d = 2$, it is well known to have an infinity of generators, with the consequence of totally killing off the degrees of freedom in any finite-$N$ constellational shape.]    
3-$d$ Shape Theory is moreover substantially less developed even in Kendall's own case of similarity shapes.   

\mbox{ }

\n{\bf Complication 2} Scaling $D$ and rotation $L$ arise as integrability conditions from the Lie bracket of a translation $P$ and a special conformal transformation $K$, 
schematically 
\be
\mbox{\bf [}P \mbox{\bf ,} \, K \mbox{\bf ]}  \m \sim \m  D + L    \m . 
\ee 
This confers a greater amount of integrability (and thus unsplittability) to the conformal group as compared to the other groups introduced so far. 

\mbox{ }

\n{\bf Complication 3} {\it Scaling} now being an integrability means that Conformal Shape Theory is a shape singleton: it has no meaningful shape-and-scale pair.

\mbox{ }

\n{\bf Remark 7} See \c{Project-1} for further discussion of Conformal Shape Theory.

\m

\n{\bf Remark 8} Affine and Conformal Geometries can be viewed as two distinct extensions of Similarity Geometry. 
In flat space, moreover, the extra generators introduced in each case are incompatible extensions: the Lie bracket interactions between these do not close.
In this way they represent a choice.
The Affine Geometry prong of this dilemma is moreover not viewed as conceptually final. 
Projective Geometry both adjoins `the point at infinity' and further conceptually simplifies affine geometry. 
For instance, Projective Geometry finishes the trivialization of the classification of conics \cite{Silvester}, 
possesses a B\'{e}zout's Theorem (in the complex case), 
projective varieties are superior to affine varieties in Algebraic Geometry \cite{Hartshorne}, 
and projective geometry includes {\sl arbitrary} observer vantage points to Image Analysis \cite{PE16}.  

\m  

\n{\bf Example 6} {\it Projective Shape Theory} \, requires passing from $\mathbb{R}^d$ carrier space to $\mathbb{RP}^{d - 1}$ carrier space 
prior to quotienting by a suitable projective group \cite{MP03, Bhatta, PE16, KKH16}.  
\be 
\FrP(N, \, d) \es \frac{\bigtimes_{i = 1}^{N}  \mathbb{RP}^{d - 1}  }{  PGL(d)}   \m .  
\ee
\n{\bf Remark 9} In general, metric-level shape(-and-scale) spaces are \cite{Kendall, GT09, Bhatta, PE16, KKH16} stratified manifolds \cite{Pflaum, BanaglBook, Kreck}.
For $d \geq 3$ metric shape-and-scale theory, the maximal coincidence-or-collision is a rather problematic separate stratum. 
For $(N, \, d) = (3, 3)$ metric shape(-and-scale) theory, the collinear configurations are a somewhat less problematic separate stratum. 
For Affine and Projective Shape Theory, however, more problematic stratification is endemic \cite{GT09, KKH16}.  

\m 

\n{\bf Example 7} For spherical carrier spaces $\mathbb{S}^d$, $d \geq 2$, the automorphisms cannot include scaling since the intrinsic curvature of the $\mathbb{S}^d$ fixes a scale. 
The Metric Shape-and-Scale Theory in this case has shape-and-scale space \cite{Kendall87, FileR, ASphe}
\be 
\FrR(N, \mathbb{S}^d) \:= \frac{\times_{i = 1}^N \mathbb{S}^d}{SO(d + 1)}                                \m . 
\ee 
\n{\bf Example 8} For toroidal carrier spaces $\mathbb{T}^d$, (including $d = 1$, for which $\mathbb{T}^1 = \mathbb{S}^1$) 
the automorphisms cannot include scaling since the topological identification involved fixes one or more scales. 
The Metric Shape-and-Scale Theory in this case has shape-and-scale space \cite{ATorus}
\be 
\FrR(N, \mathbb{T}^d)  \:=  \frac{\times_{i = 1}^N \mathbb{T}^d}{\times_{j = 1}^dU(1)} 
                       \es  \frac{\times_{k = 1}^{N \, d} \mathbb{S}^1}{\times_{j = 1}^d\mathbb{S}^1}
                       \es  \times_{l = 1}^{n \, d} \mathbb{S}^1
                       \es  \mathbb{T}^{n \, d}				                                             \m . 
\ee 
\n{\bf Remark 10} Metric Shape-and-Scale Theory on $\mathbb{S}^d$ and $\mathbb{T}^d$ thus provide two further examples of singleton theories.

\m

\n{\bf Remark 11} Let us finally consider $\Gamma$ ambiguities.
A first quartet are $\Gamma = id$, $C_2$-ref (acting reflectively), $S_N$ and $S_N \times C_2$. 
These correspond to (no, no), (yes, no), (no, yes) and (yes, yes) answers to the twofold question: do we have (mirror image, label) distinguishability? 

\m

\n{\bf Remark 12} For $N \geq 3$, moreover, partial label distinguishabilities exist, extending the above list.
Another conceptually and technically useful way \c{A-Monopoles} of viewing this extension concerns the lattice of distinguishable group actions of the subgroups of $S_N \times C_2$.
This `top group' moreover collapses to just $S_N$ if $N$ is large enough relative to $d$ 
that mirror image identification becomes obligatory by rotation through extra dimensions to those spanned by the point-or-particle separation vectors.  

\m  

\n{\bf End-Remark} As the above examples illustrate, Shape-and-Scale Theory along Kendall's lines in the generalized sense of the current section has become a rich field of study. 

\vspace{10in}

\section{Topological Shape(-and-Scale) Theories}

\n{\bf Remark 1} The current article's main focus, however, is on `rubber shapes': a coarser level of structure independent of many of the above variety of finer features. 

\m

\n{\bf Definition 1} The {\it topological notion of shape} used in this series of articles is the topological content of $N$-point constellations in dimension $d$.

\m 

\n{\bf Notation and Definition 2} Let us use pastel sky blue in depictions intended to be solely of topological shapes(-and-scales); 
I term these {\it topological distribution diagrams}.
This colouring renders them immediately distinguishable from topological shape(-and-scale) spaces (bright blue), 
                                                     geometrical shapes(-and-scales)              (grey) 
                                                  and geometrical shape(-and-scale) spaces        (black). 

\m 

\n{\bf Definition 3} {\it Coincidence diagrams} 
(see Figs \r{(1, 1)-(2, 1)-Top-Shapes}, \r{(3, 1)-Top-Shapes}, \r{(4, 1)-Top-Shapes}, \r{(3, 2)-Top-Shapes}, \r{(4, 2)-Top-Shapes} for examples) 
represent a further simplification of the previous by discarding all points not participating in coincidences.

\m 

\n{\bf Remark 2} This is a fairly weak notion of shape as many of the properties usually attributed to $N$ point constellations are metric in nature: 
the angles defining an isosceles triangle, the ratios defining a rhombus...
As the below examples show, the topological notion of shape is not however empty.
Topological shapes furthermore provide useful insights as regards the structure of shape spaces of geometric-level shapes.

\m 

\n{\bf Definition 4} These admit a considerably simpler shape(-and-scale) theory: {\it Topological Shape(-and-Scale) Theory}, characterized by just 
\be
(D, N; S, \Gamma) \m . 
\ee 
$S$ is here binary-valued: with or without scale: $\emptyset$ and $s$ standing for `scaled'. 
$D$ is ternary-valued: 1 or $\geq 1$, with the first case split furthermore into distinct open $\mathbb{R}^1$ and closed $\mathbb{S}^1$ cases; 
we denote the circle case by $1^{\prime}$ and the other two cases collectively by $d$.  

\m 

\n{\bf Remark 3} N.B.\ also what this does not depend on: neither a choice of $d \geq 2$ carrier space, 
nor a choice of continuous automorphism group aside from whether scale is included. 
The reason that scale survives as a distinction is that its inclusion necessitates the appending of the maximal coincidence-or-collision O, which remains topologically distinctive.  

\m  

\n{\bf Definition 5} The {\it topological shape space} is the collection of all of a model's topological shapes.
\be
\Top\mbox{-}\Rel(D, N; S, \Gamma) = \frac{\Top\mbox{-}\Rel(D, N; S)}{\Gamma}            \m .
\ee
These shape space graphs are to be distinguished in this article in writing by the preface $\FrT$op, and graphically by use of bright blue edges and vertices. 

\m 

\n{\bf Remark 4} Within this scheme, we can view 
\be 
{\Top\mbox{-}\Rel(D, N; S)} \m \mbox{ itself as } \m {\Top\mbox{-}\Rel(D, N; S, id)} \m .  
\ee
\n{\bf Definition 6} We also specialize to {\it topological shape spaces} \cite{A-Monopoles}
\be 
\Top\mbox{-}\FrS(N, d; \Gamma) := \Top\mbox{-}\Rel(d, N; \emptyset, \Gamma)
\ee
and {\it topological shape-and-scale spaces} \cite{A-Monopoles} 
\be 
\Top\mbox{-}\FrR(N, D; \Gamma) := \Top\mbox{-}\Rel(D, N; s \Gamma)              \m .
\ee
\n{\bf Remark 5} As we outlined in the previous section, Kendall's shape spaces are (stratified) manifolds. 
In contrast, topological shape(-and-scale) spaces are {\sl just graphs}: much simpler mathematical objects.
The vertices here are topologically-distinct configurations whereas the edges encode topological adjacency.  

\m 

\n{\bf Definition 7} The {\it complement} $\overline{\mG}$ of a graph $\mG$ has the same vertices but the complementary set of edges.

\m 

\n{\bf Remark 6} For $|\mG|$ the number of vertices of the graph $\mG$, the maximum possible number of edges is 
\be 
e(\mG)\mbox{-max}  \:=  \frac{|\mG|(|\mG| - 1)}{2}                              \m . 
\ee 
If a graph's saturation is over half of this value, it is usually more straightforward to characterize, recognize and depict in terms of its complement.  

\m 

\n{\bf Definition 7} The {\it cone graph}, $\mC(\mG)$, of a given graph $\mG$ has all of $\mG$'s edges and vertices plus one vertex which is joined by |\mG| further edges, 
one to each vertex of $\mG$.  

\m 

\n{\bf Lemma 1}  If $\mG$ is a given topological shape theory's shape space, 
then the cone graph $\mC(\mG)$ is the corresponding topological shape-and-scale theory's shape-and-scale space. 
The cone vertex here is the maximal coincidence-or-collision O. 

\m 

\n{\bf Lemma 2} Cone graphs admit a trivial characterization in terms of graph complements, 
\be 
\overline{\mC(\mG)} = \overline{\mG} \coprod \mD_1  \m ,
\ee 
where $\mD_1$ is the singleton graph (discrete graph of order 1, alias complete graph of order 1, $\mK_1$ and path graph of length 1, $\mP_1$). 

\m 

\n{\bf Remark 7} Cone graphs also feature as shape spaces fof $d \geq 2$ because in these cases the generic configuration $\mG$ is topologically adjacent to 
all other configurations.  

\m 

\n{\bf Remark 8} Cones over cones of graphs, $\mC(\mC(\mG))$, also feature in the corresponding $d \geq 2$ shape-and-scale spaces, with O and G as cone points 
(in either order).
Because of this, the following Corollary is also used. 

\m 

\n{\bf Corollary 1} Cone of a cone graphs also admit a trivial characterization in terms of graph complements, 
\be 
\overline{\mC(\mC(\mG))} = \overline{\mG} \coprod \mD_1 \coprod \mD_1  \m .  
\ee 
\n{\bf Definition 7} The {\it suspension graph}, $\mS(\mG)$, of a given graph $\mG$ has all of $\mG$'s edges and vertices plus two vertices which are each joined by |\mG| 
further edges, one to each vertex of $\mG$.  

\m 

\n{\bf Remark 9} These are not the same as cones over cones, since the two suspension points are not themselves joined by an edge, giving the following characterization. 

\m 

\n{\bf Lemma 3} Suspension graphs admit a trivial characterization in terms of graph complements, 
\be 
\overline{\mC(\mG)} = \overline{\mG} \coprod \mP_2                      \m ,
\ee 
where $\mP_2$ is the 2-path graph (alias complete graph of order 2, $\mK_2$). 

\m 

\n{\bf End-Remark}  Euclidean and Equi-top-voluminal Shape-and-Scale Theories' scale  is a mathematically trivial appendage at the group-theoretic level.   
By Lemma 2, scale is a trivial appendage at the level of topological graphs as well. 
Including scale is not however trivial at the metric level, firstly on topological grounds: for instance spheres and real spaces are homotopically distinct. 
Secondly, on stratificational grounds, since O is a separate stratum.  
By these considerations, and Complication 3 and Remark 10 of the previous section, 
scale being regarded as a trivial appendage becomes highly contextual and dubious, even just on mathematical grounds.  
 
\vspace{10in}
 
\section{(1, 1), (2, 1) and (3, 1) examples}\l{(1,1)-(2,1)}

\n For clarity, we state that the carrier space in the current section is the open case of connected, boundary-less 1-$d$ manifold: $\mathbb{R}^1$  

\m 

\n{\bf Example 1} For $N = 1$, there is one topological class: the point-or-particle itself (Fig \r{(1, 1)-(2, 1)-Top-Shapes}.1).    

\m 

\n For $N = 1$, the maximal coincidence-or-collision notion coincides with the point-or-particle notion itself. 

\m 

\n For arbitrary $(N, \, d)$, removing $O$ from consideration restricts one to the normalizable shapes (i.e.\ of finite total moment of inertia). 

\m 

\n For $N = 1$, however, there are exceptionally no normalizable shapes, so the latter modelling consideration leaves on with just the empty set.  

\m

\n{\bf Remark 1} Moreover, even including the O point, this model exhibits other trivialities
(a translation-invariant single-particle universe model has no content by Leibniz's Identity of Indiscernibles), so one passes to considering $N \geqs 2$.
However, even including O the model is still metric dimensionally trivial, giving plenty of reasons to pass to $N \geqs 3$.
%
{            \begin{figure}[!ht]
\centering
\includegraphics[width=0.42\textwidth]{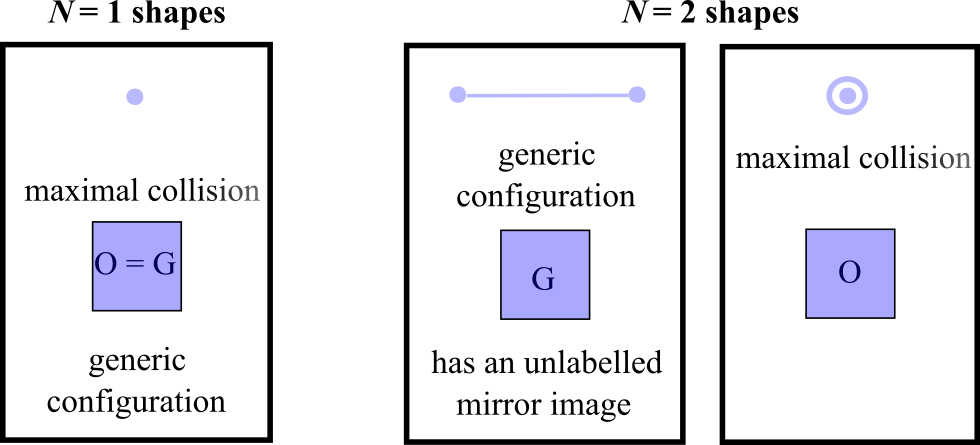}
\caption[Text der im Bilderverzeichnis auftaucht]{        \footnotesize{Topological classes of configurations for 1 and 2 points.  
O denotes the maximal coincidence-or-collision and G denotes the coincidence-or-collision-less generic configuration.  
The 2 point case has distinct classes because, while topology has no notion of length (`rubber sheet'), 
it does distinguish between coincidence and non-coincidence: passing from a coincidence to a non-coincidence involves a tearing and the reverse passage a gluing.}}
\l{(1, 1)-(2, 1)-Top-Shapes} \end{figure}          }

\m 

\n{\bf Example 2} For $N = 2$, there are two topological classes of (non-)shape (Fig \r{(1, 1)-(2, 1)-Top-Shapes}.2). 

\m

\n In this case, excluding O still leaves us with a topological theory.

\m 

\n Moreover, excluding O from $N = 2$ leaves one with no topological class distinction; one needs to consider at least $N = 3$ to have this feature.

\m 

\n Many reasons for excluding O remain absent for $N = 2$ as well.
All in all, the binary coincidence-or-collision's good behaviour turns out to trump the maximal coincidence-or-collision's bad behaviour 
in this case in which these two notions coincide.  

\m 

\n{\bf Example 3} and {\bf Proposition 1} For $N = 3$, there are three topological classes of (non-)shape: Fig \r{(3, 1)-Top-Shapes}.

\m 

\n{\bf Remark 2} This is the first dimensionally-nontrivial metric shape space, for all that it is still relationally trivial as a metric shape space:  
it has 1 degree of freedom, whereas relational theories require at least two degrees of freedom so that one can change relative to the other.  
The corresponding shape-and-scale relational space, moreover, is relationally nontrivial out of having 2 degrees of freedom.  
%
{            \begin{figure}[!ht]
\centering
\includegraphics[width=0.35\textwidth]{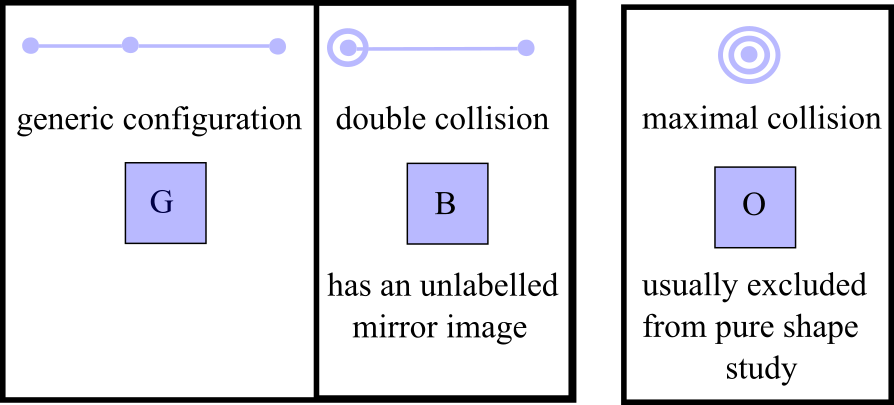}
\caption[Text der im Bilderverzeichnis auftaucht]{        \footnotesize{Topological classes of configurations for 3 points in 1-$d$.  
Here B denotes the binary coincidence-or-collision B, and G denotes the coincidence-or-collision-less generic shape.} }
\l{(3, 1)-Top-Shapes} \end{figure}          }

\m 

\n{\bf Proposition 2} These topological classes of shapes may furthermore be viewed as {\sl equivalence classes} of metric shapes. 

\m 

\n{\bf Remark 4} Equivalence classes are always disjoint and exhaustive and so constitute a partition.  
This is the current article's coarsest-level characterization of topological shape (or, for that matter, of shape).  

\m 

\n{\bf Proposition 3} For $N = 3)$, the number and types of topological classes themselves are a dimension $d$-independent characterization.  

\m 

\n{\bf Remark 5} Which metric shape representations each class contains is moreover dimension-dependent: compare Fig \r{(3, 1)-Top-Shapes} with Fig \r{(3, 2)-Top-Shapes}.   

\m 

\n{\bf Proposition 4} If one or both of point-or-particle labelling and mirror image distinction are incorporated, a finer decomposition of the topological shapes is being entertained.  

\m  

\n{\bf Example 1 Revisited}  For $N = 1$, both of these distinctions are meaningless, so there is still just one point-or-particle.

\m 

\n{\bf Example 2 Revisited}  For $N = 2$, these distinctions are equivalent, 
in the sense that `left' and `right' assignment has the same labelling content as any other specification of 2 distinct labels. 
In both cases, there are now 2 G's in place of 1, and still just room for 1 realization of $\mB = \mO$.

\m 

\n{\bf Example 3 Revisited} For $N = 3$, if the points-or-particles are labelled and mirror image shapes are held to be distinct, 
\be
\mbox{\#(G)} = \mbox{(label permutations)} = 3 \, ! = 6\mma \mbox{ and }
\ee
\be 
\mbox{\#(B)} = \mbox{(3 ways of leaving a particle out)} \times \mbox{(2 mirror images)} = 6 \m  .  
\ee
Suppose instead that the points are labelled but mirror image shapes are held to be identified.
Then these values are halved by the mirror image identification to 3 of each.
If the points are moreover not labelled but mirror images are held to be distinct, 2 distinct G's and 2 distinct B's ensue.
Finally, if neither distinction is made, there is just 1 of each.  

\m 

\n{\bf Remark 5} The distinction between `equivalence classes in general give partitions' 
and the unlabelled mirror image identified topological shape-and-scale configurations for (3, 1) is in 1 : 1 correspondence with {\sl the} finite partitions themselves 
(row 3 of Fig \r{Top-Shape-Coincidence}).  

\m 

\n Sec 5 shows moreover that this 1 : 1 correspondence is {\sl not} maintained for $N \geqs 4$; instead a somewhat finer partition is realized. 
On the other hand, Sec 7 shows that $d \geqs 2$ also maintains this correspondence for arbitrary $N$. 
This motivates introducing a second type of depiction which keeps track of {\sl the} finite partition content, as follows, 
for all that for the (3, 1) model these two types have equivalent content.   
%
{            \begin{figure}[!ht]
\centering
\includegraphics[width=0.65\textwidth]{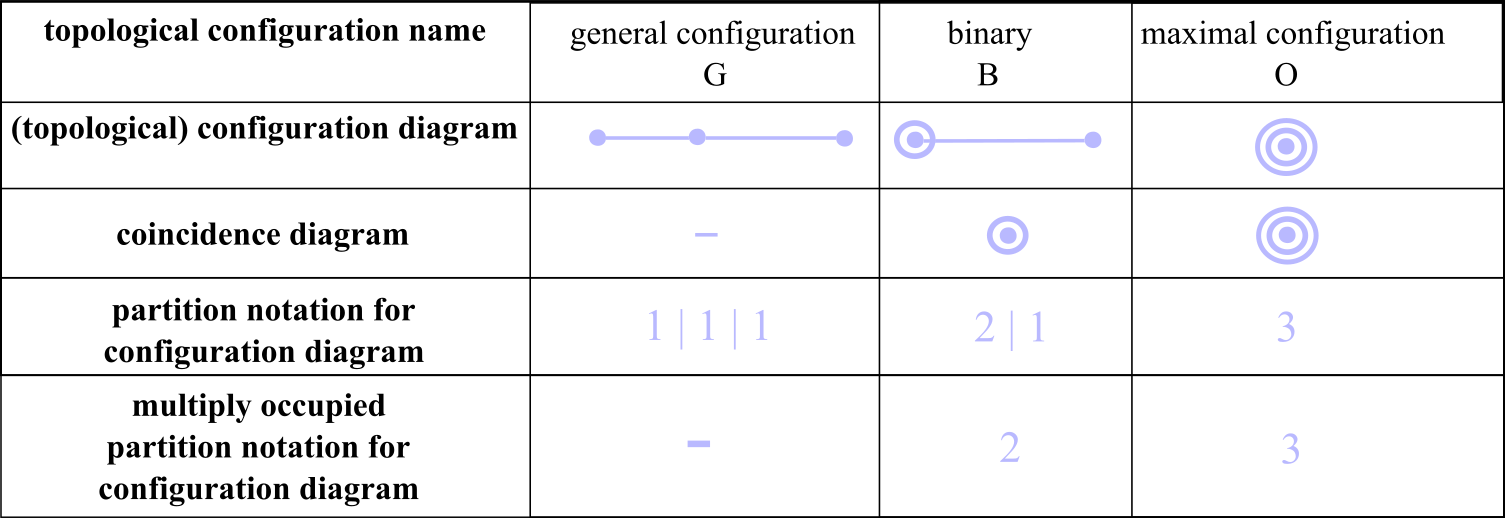}
\caption[Text der im Bilderverzeichnis auftaucht]{        \footnotesize{Topological shape diagrams versus coincidence diagrams, with some corresponding partition notation. } }
\l{Top-Shape-Coincidence} \end{figure}          }

\m   

\n{\bf Proposition 5} For $N = 1$, i) excluding O the sole topological shape space is
\be
\Top\mbox{-}\FrS(1, 1)          := 
\Top\mbox{-}\FrS(1, 1; id)       =  \emptyset \m  .  
\ee
\n ii) Including O -- which we indicate by appending an O superscript to the shape space in question --
\be
\Top\mbox{-}\FrS^{\sO}(1, 1)             := 
\Top\mbox{-}\FrS\mbox{}^{\sO}(1, 1; id)   =  \mP_1: \mbox{  a single point labelled with } \mO = \mG        \m  .  
\ee
\n{\bf Proposition 6} For $N = 2$, i) excluding O, the topological shape spaces are
\be
\Top\mbox{-}\FrS(2, 1)  :=  \Top\mbox{-}\FrS(2, 1)        
                         =  P_1 \, \coprod \, P_1 \mbox{ : two points labelled by } \mG                       \m  . 
\l{Top-S(2, 1)}
\ee
Also 
\be
\Top\mbox{-}\FrS(2, 1; C_2)      =  
\Top\mbox{-}\Leib_{\sFrS}(2, 1)  =  \mP_1   \mbox{ : a single point labelled with } \mG                      \m  .  
\ee
ii) Including O, 
\be
\Top\mbox{-}\FrS^{\sO}(2, 1)      := 
\Top\mbox{-}\FrS^{\sO}(2, 1; id)   =   \mP_3                                                                \m  :
\ee
the 3-vertex {\it 3-path graph} labelled as per Fig \r{S(1, 1)-S(2, 2)-Top}.c.1).  
Also 
\be
\Top\mbox{-}\Leib_{\sFrS}^{\sO}(2, 1)  = 
\Top\mbox{-}\FrS^{\sO}(2, 1; C_2)      =   \mP_2                                                            \m  :  
\ee
the 2-vertex {\it 2-path graph} labelled as per Fig \r{S(1, 1)-S(2, 2)-Top}.c.2).  
%
{            \begin{figure}[!ht]
\centering
\includegraphics[width=0.72\textwidth]{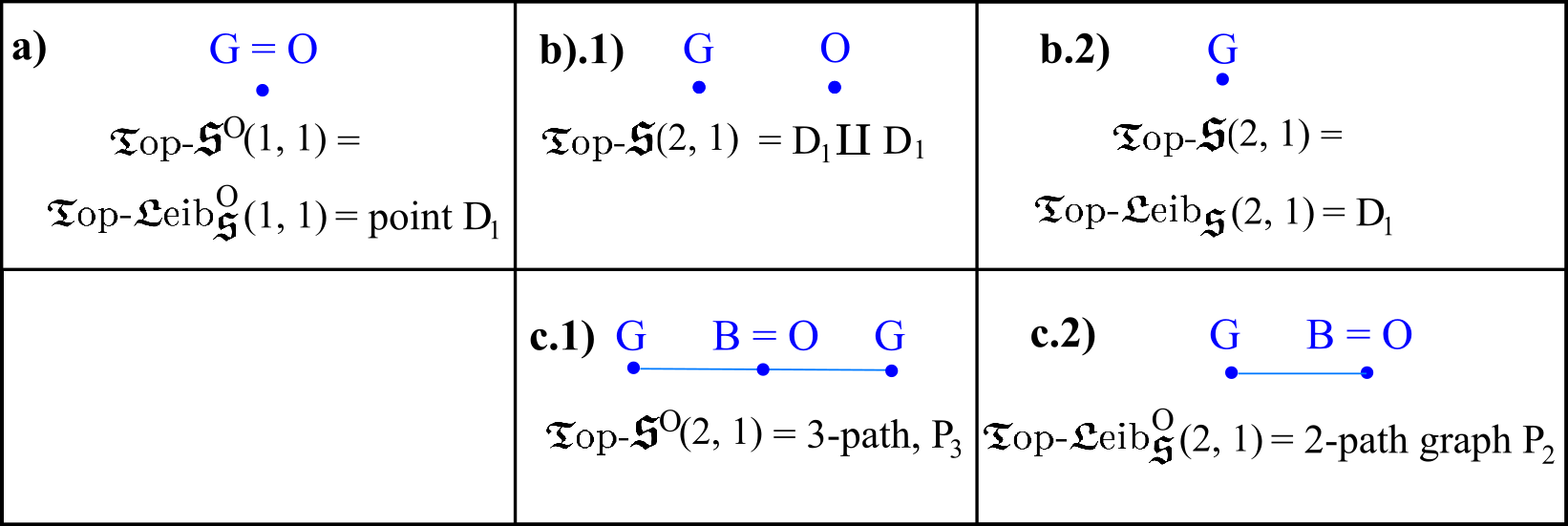}
\caption[Text der im Bilderverzeichnis auftaucht]{        \footnotesize{The smallest topological shape(-and-scale) spaces.    } }
\l{S(1, 1)-S(2, 2)-Top} \end{figure}          }

\m 

\n{\bf Remark 6} (\r{Top-S(2, 1)}) is disconnected: there is no edge between the vertices of this graph.  
On the other hand, the structure of $\Top\mbox{-}\FrS^{\sO}(2, 1)$ can be arrived at by a continuity method (Fig \r{S(1, 1)-S(2, 2)-Top}.c.0).   

\m 

\n{\bf Proposition 7} For $(N, \, d) = (3, 1)$, the maximal topological shape space is  
\be
\Top\mbox{-}\FrS(3, 1)      \:=  
\Top\mbox{-}\FrS(3, 1; id)  \es  \mC_{12}   \m :
\ee
the 12-vertex cycle graph as labelled in Fig \r{S(3, 1)-Top}.1).     

\m 

\n{\bf Remark 7)} In topological shape(-and-scale) graphs, the shapes are just vertex labels.
On the other hand, the graph edges encode the {\it topological adjacency} relation.   
This is based on topology distinguishing between plain tearing and tearing followed by gluing to another distinguishable object. 
So e.g.\ AB--C is not topologically adjacent to A--BC, since to move between these, one needs to tear B off A and then glue it to C.   

\m

\n{\bf Proposition 8} For $(N, \, d) = (3, 1)$, the lattice of subgroup actions is as per Fig \r{(3, 1)-Latt}.  
%
{            \begin{figure}[!ht]
\centering
\includegraphics[width=0.21\textwidth]{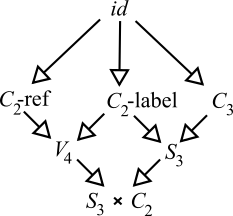}
\caption[Text der im Bilderverzeichnis auftaucht]{        \footnotesize{Lattice of distinct subgroup actions of $\sFrS(3, 1)$.} }
\l{(3, 1)-Latt} \end{figure}          }

\m 

\n{\bf Proposition 10} For $(N, \, d) = (3, 1)$, the other topological shape spaces are as follows.  
\be 
\Top\mbox{-}\FrS(3, 1; C_2\mbox{-ref})  \:=  \frac{\Top\mbox{-}\FrS(3, 1)}{C_2\mbox{-ref}} 
                                        \es  \mC_{6}                                               \m :
\ee
the 6-vertex cycle graph as labelled in Fig \r{S(3, 1)-Top}.2). 
\be
\Top\mbox{-}{\FrS}(3, 1; C_2\mbox{-label})  \:=  \frac{\Top\mbox{-}\FrS(3, 1)}{C_2\mbox{-label}} 
                                            \es  \mP_{7}                                           \m : 
\ee 
the 7-vertex path graph as labelled in Fig \r{S(3, 1)-Top}.3). 
\be
\Top\mbox{-}{\FrS}(3, 1; C_3)  \:=  \frac{\Top\mbox{-}\FrS(3, 1)}{C_3} 
                               \es  \mC_{4}                                                        \m : 
\ee 
the 4-vertex cycle graph as labelled in Fig \r{S(3, 1)-Top}.4). 
\be
\Top\mbox{-}{\FrS}(3, 1; V_4)  \:=  \frac{\Top\mbox{-}\FrS(3, 1)}{V_4} 
                               \es  \mP_{4}                                                        \m : 
\ee 
the 4-vertex path graph as labelled in Fig \r{S(3, 1)-Top}.5). 
\be
\Top\mbox{-}\FrI\FrS(3, 1; S_3)  \:=  \frac{\Top\mbox{-}\FrS(3, 1)}{S_3} 
                                 \es  \mP_{3}                                                      \m  : 
\ee
the 3-vertex path graph with the symmetric endpoint labelling given in Fig \r{S(3, 1)-Top}.6).  
\be
\Top\mbox{-}\Leib_{\sFrS}(3, 1)  \:=  \Top\mbox{-}\FrI\FrS(3, 1; S_3 \times C_2)
                                 \es  \frac{\Top\mbox{-}\FrS(3, 1)}{S_3 \times C_2} 
								 \es   \mP_2                                                       \m  :
\l{Top-Leib(3, 1)}
\ee
the 2-vertex path graph with end-points labelled distinctly as per Fig \r{S(3, 1)-Top}.7). 
%
{            \begin{figure}[!ht]
\centering
\includegraphics[width=0.62\textwidth]{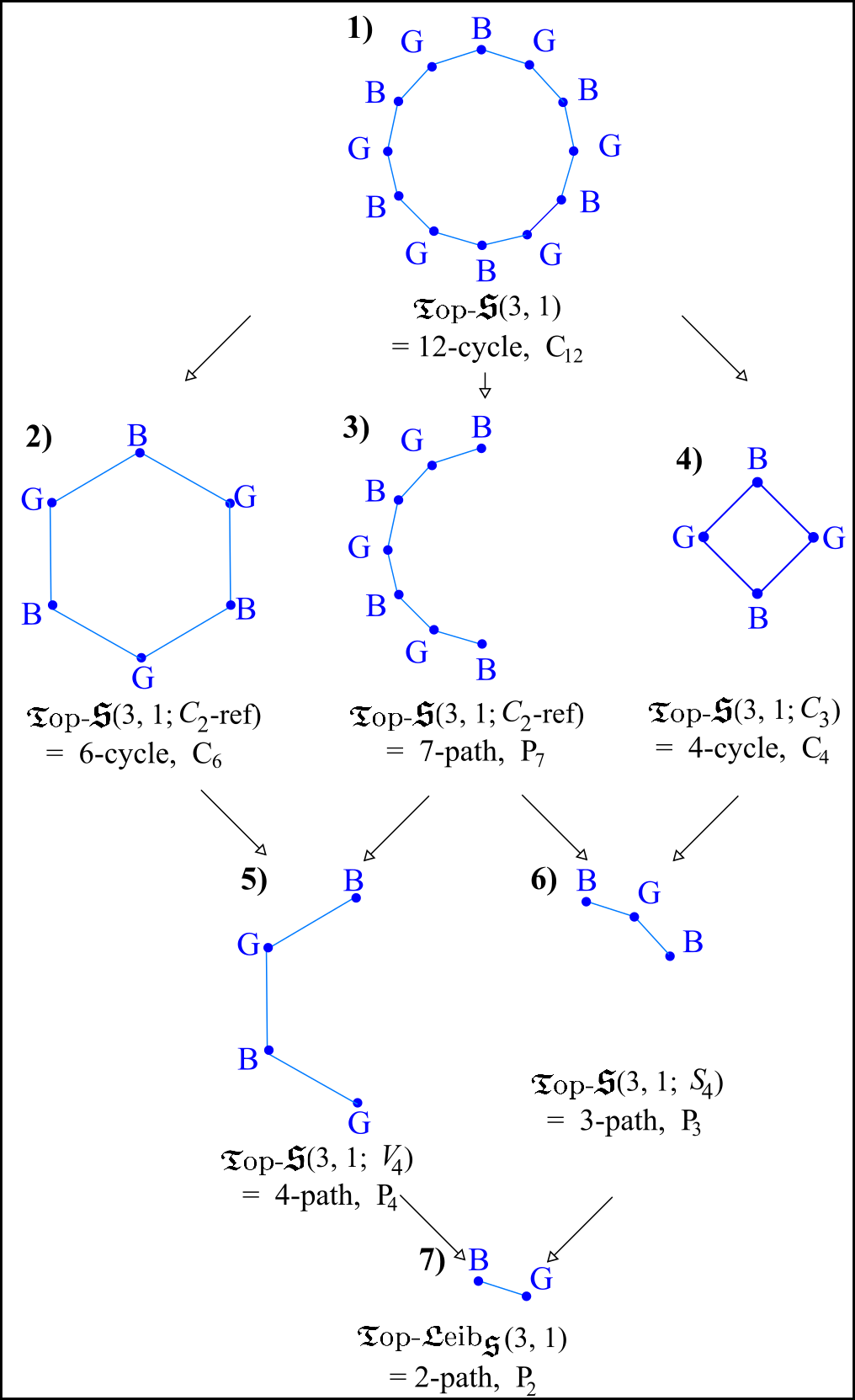}
\caption[Text der im Bilderverzeichnis auftaucht]{        \footnotesize{(3, 1) topological shape spaces.} }
\l{S(3, 1)-Top} \end{figure}          }

\m 

\n {\bf Corollary 1} The B's or G's of $\Top\mbox{-}\Leib_{\sFrS}(3, 1)$ form a hexagon, and  
 \be
S_3 \times C_2 \m  \mbox{ acts on this hexagon as } \m  D_6 \mbox{ : } \mbox{ the dihaedral group of order 12 } .
\ee
This permits us to rewrite the labelling and mirror image based definition (\r{Top-Leib(3, 1)}) as 
\be
\Top\mbox{-}\Leib_{\sFrS}(3, 1)  \es  \frac{\Top\mbox{-}\FrS(3, 1)}{D_6}\m  ,
\ee
which is `more natural' at the shape space level because of the realization of the hexagons therein.

\m 

\n{\bf Structure 1} In shape(-and-scale) space differential geometries,   
it is helpful and structurally meaningful to draw the shapes each point represents as follows 
(the start of a standardized and mathematically precise way of drawing Kendall's spherical blackboard and generalizations).  

\m 

\n{\bf Proposition 10} For For $(N, \, d) = (3, 1)$, the maximal topological shape-and-scale space is  
\be
\Top\mbox{-}\FrR(3, 1)      := 
\Top\mbox{-}\FrR(3, 1; id)   = \mW_{12}                                                              \m :
\ee
the 13-vertex 12-spoked {\it wheel graph} \c{I} as labelled in Fig \r{R(3, 1)-Top}.1).     

\m

\n{\bf Proposition 11} For $(N, \, d) = (3, 1)$, the other topological shape-and-scale spaces are then as follows.  
\be 
\Top\mbox{-}\FrR(3, 1; C_2\mbox{-ref}) \:=  \frac{\Top\mbox{-}\FrR(3, 1)}{C_2\mbox{-ref}} 
                                        \es   \mW_{6}                                                \m :
\ee
the 7-vertex 6-spoked wheel graph as labelled in Fig \r{R(3, 1)-Top}.2). 
\be
\Top\mbox{-}{\FrR}(3, 1; C_2\mbox{-label}) \:=  \frac{\Top\mbox{-}\FrR(3, 1)}{C_2\mbox{-label}} 
                                           \es  \mF_{6}                                              \m : 
\ee 
the 8-vertex 6-panelled fan graph as labelled in Fig \r{R(3, 1)-Top}.3). 
\be
\Top\mbox{-}{\FrR}(3, 1; C_3)  \:=  \frac{\Top\mbox{-}\FrR(3, 1)}{C_3} 
                               \es  \mW_{4}                                                          \m : 
\ee 
the 5-vertex 4-spoked wheel graph as labelled in Fig \r{R(3, 1)-Top}.4). 
\be
\Top\mbox{-}{\FrR}(3, 1; V_4)  \:=  \frac{\Top\mbox{-}\FrR(3, 1)}{V_4} 
                               \es  \mF_{3}  
                                =   \mbox{gem}					                                     \m : 
\ee 
the 5-vertex 3-panelled fan graph alias gem graph as labelled in Fig \r{S(3, 1)-Top}.5). 
\be
\Top\mbox{-}\FrR(3, 1; S_3)  \:=  \frac{\Top\mbox{-}\FrR(3, 1)}{S_3}  
                                 \es  \mF_{2}
							      =   \mbox{diamond}                                                 \m  : 
\ee
the 4-vertex 2-panelled fan graph alias diamond graph with the symmetric endpoint labelling given in Fig \r{R(3, 1)-Top}.6).  
\be
\Top\mbox{-}\Leib_{\sFrR}(3, 1)                  \:=  
\Top\mbox{-}\FrI\FrR(3, 1; S_3 \times C_2)       \es  \frac{\Top\mbox{-}\FrR(3, 1)}{S_3 \times C_2} 
                                                  =   \mC_3                                          \m  :
\l{Top-Leib-R(3, 1)}
\ee
the 3-vertex cycle graph labelled as per Fig \r{R(3, 1)-Top}.7). 
%
{            \begin{figure}[!ht]
\centering
\includegraphics[width=0.4\textwidth]{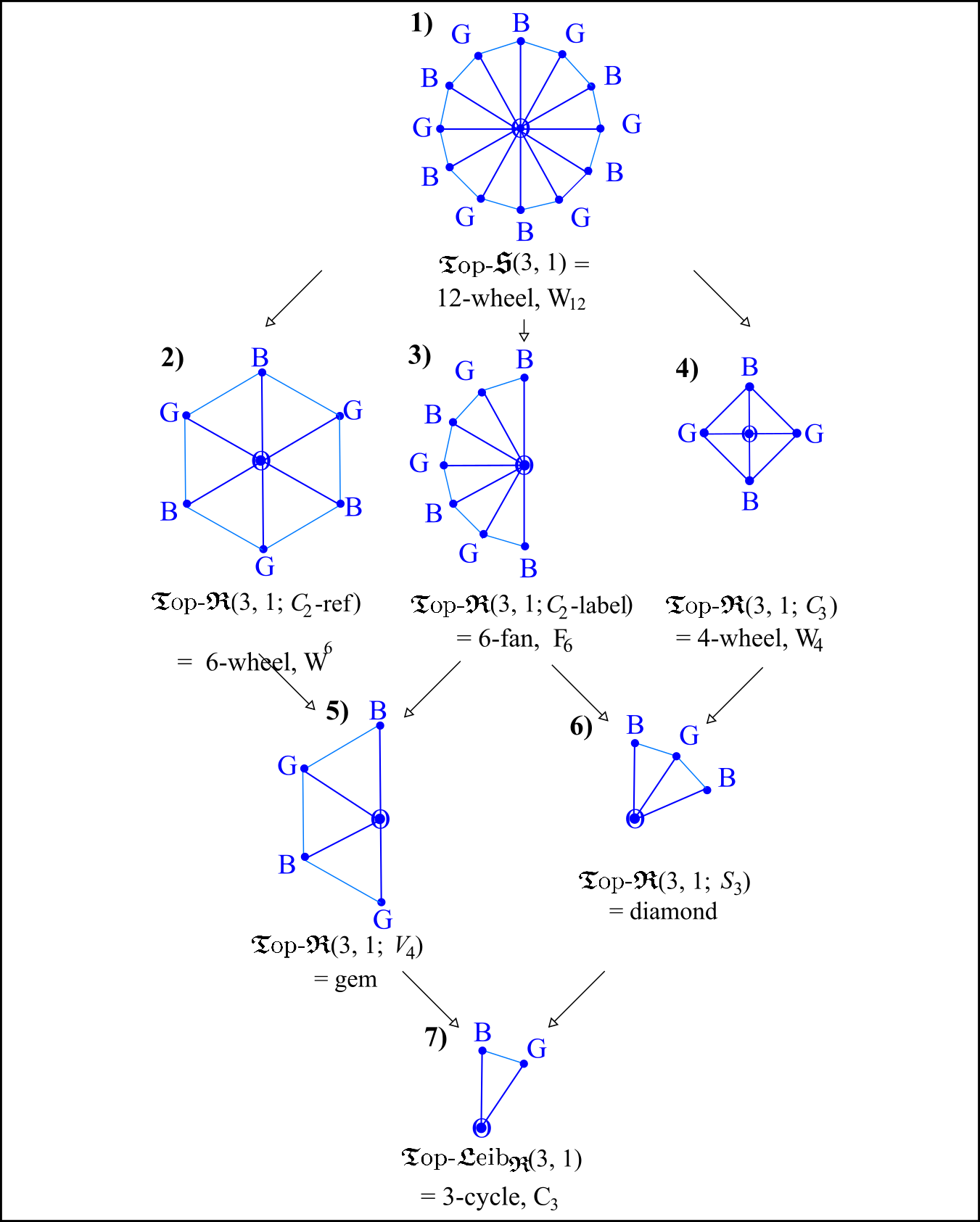}
\caption[Text der im Bilderverzeichnis auftaucht]{        \footnotesize{(3, 1) topological shape-and-scale spaces.} }
\l{R(3, 1)-Top} \end{figure}          }

\vspace{10in}

\section{(1$^{\prime}$, 1), (2$^{\prime}$, 1) and (3$^{\prime}$, 1) examples}\l{S1-Examples}

\n The carrier space is now $\mathbb{S}^1$  
This model possesses no shape spaces because there is no overall dilation generator.  
The $N = 1$ and $N = 2$ topologically shape-and-scale spaces are as per the previous section.
The $N = 3$ case is moreover different, as follows. 

\m

\n{\bf Proposition 1} The distinct $(\mathbb{S}^1, 3)$ topological shape-and-scale spaces are the cones over the graphs in Fig \ref{S1} with O as cone point. 
Sweeping down and right, Lemma 2 of Sec 3 gives $\overline{\mK_3 \coprod \mK_2 \coprod K_1}$, $\overline{\mK_3 \coprod \mK_1 \coprod K_1}$, two labellings of the diamond graph, and $\mC_3$.  
%
{            \begin{figure}[!ht]
\centering
\includegraphics[width=0.36\textwidth]{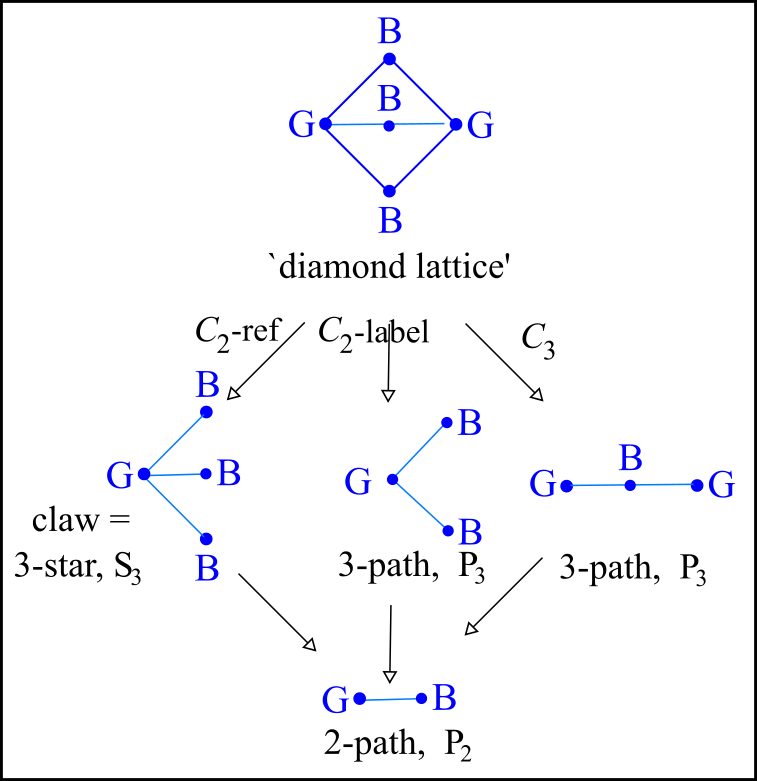}
\caption[Text der im Bilderverzeichnis auftaucht]{        \footnotesize{Graphs used in characterizing $(\mathbb{S}^1, 3)$ topological shape-and-scale spaces.
Note that the `diamond lattice' is a suspension graph with suspension points G and G.} }
\label{S1} \end{figure}          }

\vspace{10in}

\section{(4, 1) Examples}\l{(4,1)}

{\bf Proposition 1} There are 6 topological types of (4, 1) shape, as per Fig \ref{(4, 1)-Top-Shapes}.
%
{            \begin{figure}[!ht]
\centering
\includegraphics[width=0.8\textwidth]{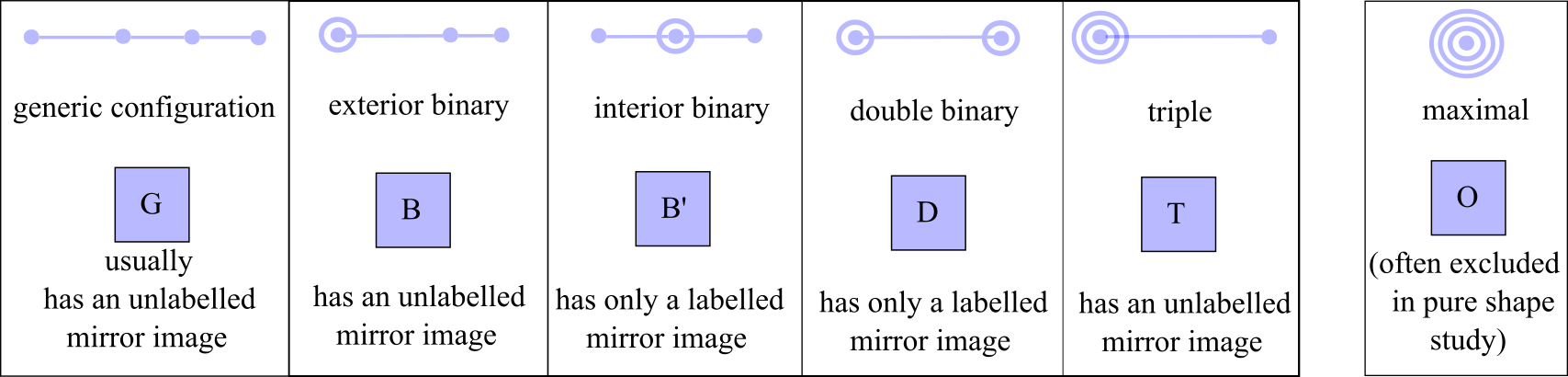}
\caption[Text der im Bilderverzeichnis auftaucht]{        \footnotesize{The 6 topological types of (4, 1) shape: 
generic configuration G, 
exterior and interior binary coincidences-or-collisions denoted by B and B$^{\prime}$ respectively, 
double binary coincidences-or-collisions D, 
ternary coincidences-or-collisions T, 
and the usually excluded maximal coincidence-or-collision O.} }
\label{(4, 1)-Top-Shapes} \end{figure}          }

\m

\n{\bf Remark 1} This is the smallest example in which 1-$d$ mirror-image-identified topological shapes are not just partitions; $N \leq 3$ has just partitions. 
For $N = 4$ the                            $2 \, | \, 1 \, | \, 1$ partition is further fine-grained by the binary coincidences-or-collisions being split into 
the exterior     $\mB$'s                   $1 \, | \, 2 \, | \, 1$ 
and the interior $\mB^{\prime}$'s          $2 \, | \,1 \, | \, 1$. 
This corresponds to topological shapes in 1-$d$ possessing an additional notion of order in which partitions are realized, and is further explained as Corollary 2 in Sec 8.   

\m

\n{\bf Remark 2} If mirror images are distinct,      the $2 \, | \, 1 \, | \, 1$ partition 
                                    is distinct from the $1 \, | \, 1 \, | \, 2$ as well.  
This is not a new effect, however, since (3, 1) already suffices to split the binary coincidences-or-collisions into $2 \, | \, 1$ and 
                                                                                                                 and $1 \, | \, 2$ partitions.  

\m

\n{\bf Remark 3} For labelled and mirror image distinct (4, 1) shapes,  
\be
\#(\mG) = \left( \mbox{label permutations } \right) = 4 \, ! 
        = 24                                                                                                       \m .
\ee
\be 
\#(\mB) = \left( C(4, 2) \mbox{ choices of pair } \right)       \times 
          \left(  \mbox{ 2 orders for other particles } \right) \times 
		  \left( \mbox{ 2 mirror images } \right) 
        = 24                                                                                                       \m . 
\ee
\be 
\#(\mB^{\prime}) = \left( C(4, 2) \mbox{ choices of pair} \right) \times \left(  \mbox{ 2 orders for other particles or 2 mirror images }  \right) 
                 = 12                                                                                                                                               
\ee
since now these two doubling effects are coincident, rather than cumulative, by the topologically symmetrical central positioning of the binary coincidence-or-collision.
\be 
\#(\mD) = \left( \m C(4, 2) \mbox{ choices of pair } \right) 
        = 6                                                                                                         \m . 
\ee
\be 
\#(\mT) = \left( \mbox{ 4 ways of leaving 1 particle out } \right) \times \left( \mbox{ 2 mirror images } \right) 
        = 8                                                                                                         \m .  
\ee
\n{\bf Proposition 1} For mirror images held to be distinct and distinguishably labelled points, the topological shape space is
\be
\Top\mbox{-}\FrS(4, 1) = ( \mbox{ 74-vertex cube graph } )                                                          \m ,
\ee
labelled as in Fig \r{S(4, 1)-Top-beta}.1).     

\m

\n{\u{Derivation}}. Continuity considerations show that these shapes fit together in the manner of Fig \r{S(4, 1)-Top-gamma}, which closes up to form Fig \r{S(4, 1)-Top-beta}.1).  
See Fig \r{(4, 1)-Graphs}.1) for a full explicit graphic representation (as a planar graph).
%
{            \begin{figure}[!ht]
\centering
\includegraphics[width=0.5\textwidth]{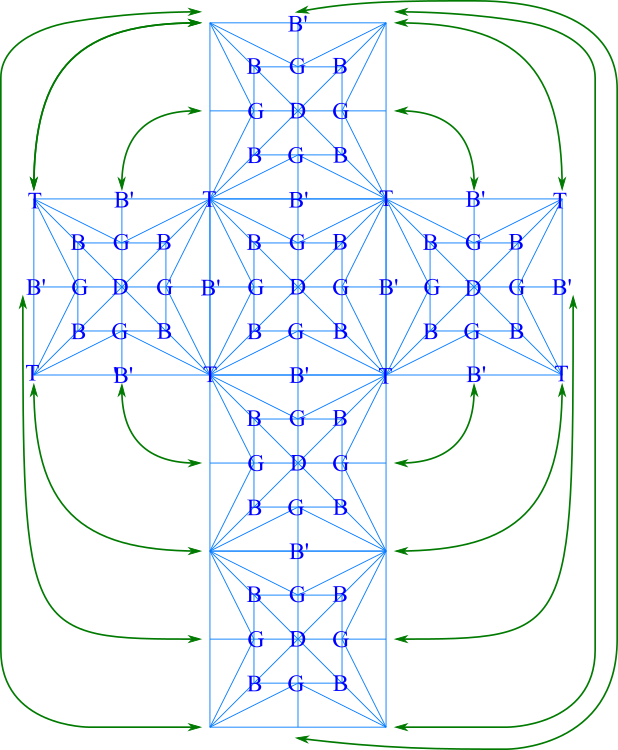}
\caption[Text der im Bilderverzeichnis auftaucht]{        \footnotesize{Continuity method for determining the topology of $\Top\mbox{-}\sFrS(4, 1)$; 
the green arrows indicate topological identification.} }
\label{S(4, 1)-Top-gamma} \end{figure}          }

\m 

\n{\bf Remark 4} For $(N, \, d) = (4, 1)$, the maximal discrete group is $S_4 \times C_2$. 
On the one hand, this is of order 48, which is larger than the previous section's order 12, and supports more divisors and subgroups. 
On the other hand, the object acted upon is the 74-vertex cube graph: the largest graph in the current article. 
By this combination of complexities, the number of distinct subgroup actions is sizeable, and the graphs thus produced are complicated. 
Because of this, we do not provide the whole lattice of graphs for the (4, 1) shape spaces. 
Instead we consider another of the largest examples from near the top of the lattice and four of the smallest examples from around the bottom of the lattice.  
%
{            \begin{figure}[!ht]
\centering
\includegraphics[width=0.7\textwidth]{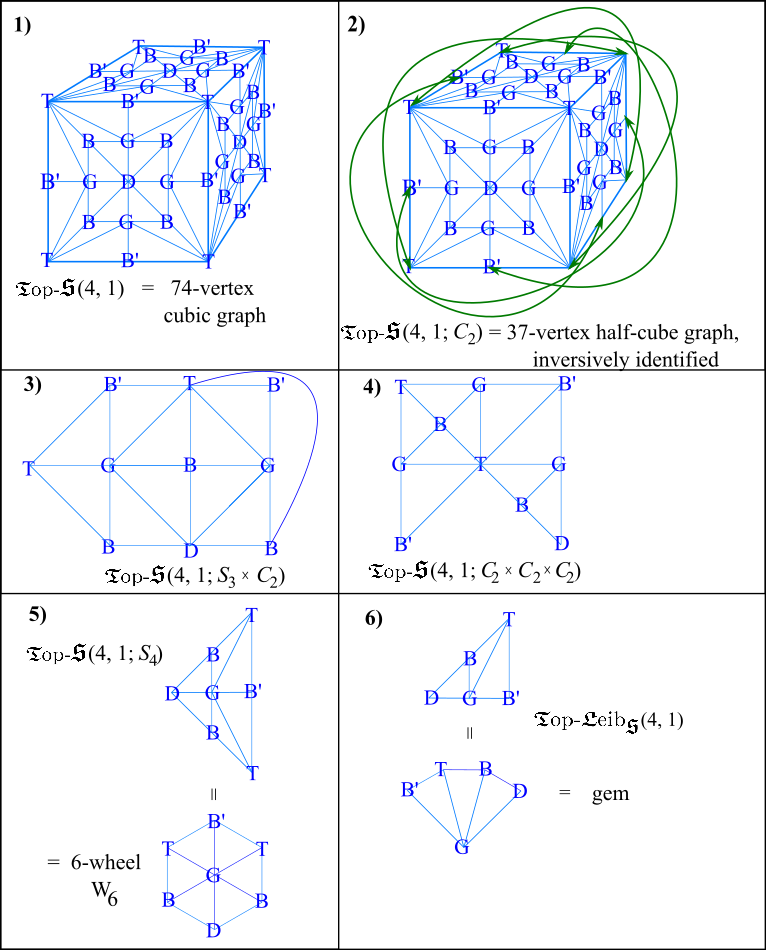}
\caption[Text der im Bilderverzeichnis auftaucht]{        \footnotesize{(4, 1) topological shape spaces. } }
\label{S(4, 1)-Top-beta} \end{figure}          }

\m 
 
\n{\bf Proposition 3} If mirror images are identified but labels remain distinct, the topological shape space is 
\be 
\Top\mbox{-}{\FrS}(4, 1; C_2\mbox{-ref})   \:=  \frac{  \Top\mbox{-}\FrS(4, 1)  }{  C_2\mbox{-ref}  } 
                                           \es  ( \mbox{ 37-point inversively-identified half-cube graph } )         \m ,  
\ee
as labelled in Fig \ref{S(4, 1)-Top-beta}.2), and fully depicted as a graph in Fig \ref{(4, 1)-Graphs}.2).   

\m

\n{\bf Proposition 4} For indistinguishable labels with mirror images identified, the topological Leibniz shape space -- the bottom element of the lattice -- is 
\be
\Top\mbox{-}\Leib_{\sFrS}(4, 1)  := 
\Top\mbox{-}\FrS(4, 1; S_4 \times C_2) \:=  \frac{\Top\mbox{-}\FrS(4, 1)}{S_4 \times C_2} 
                                       \es  \mbox{gem}  
                                        =	\m \mbox{ 3-fan } \m  \mF_3						\m :
\label{Leib(4, 1)}
\ee
with labels all distinct as per Fig \r{S(4, 1)-Top-beta}.6).

\m    

\n{\bf Proposition 5} If indistinguishable labels are considered instead while mirror image distinction is retained, the topological shape space is 
\be
\Top\mbox{-}\FrS(4, 1; S_4) \:=  \frac{\Top\mbox{-}\FrS(4, 1)}{S_4} 
                            \es  \mW_6                                                                               \m : 
\ee
the 6-spoked wheel graph labelled as per Fig \r{S(4, 1)-Top-beta}.5).

\m

\n{\bf Proposition 6} If partially distinguishable labels of the form AAAB are considered alongside mirror image identification, the topological shape space is 
\be
\Top\mbox{-}\FrS(4, 1; S_3 \times C_2) = \mbox{ 10-vertex 18-edge graph of Fig \ref{S(4, 1)-Top-beta}.3) } . 
\ee 
This is re-represented in \ref{(4, 1)-Graphs}.3) as a rectilinear planar graph.   

\m  

\n{\bf Proposition 7} Consider partially distinguishable labels of the form AABB with A and B additionally {\it meaningless} 
so swapping A and B furthermore makes no difference.\footnote{Quark colours are a well-known example of a triplet of labels that are meaningless in this sense}. 
Suppose furthermore that mirror image identification applies.  
Then the topological shape space is 
\be
\Top\mbox{-}\FrS(4, 1; C_2 \times C_2 \times C_2) = \mbox{ 10-vertex 18-edge graph of Fig \ref{S(4, 1)-Top-beta}.4) } . 
\ee
\n{\bf Remark 5} The previous two graphs are clearly non-isomorphic at the level of topological shape graph labels, 
since the first has 2 G's, 3 B's, 2 T's and 1 D, whereas the second has 3 G's, 2 B's, 1 T and 2 D's. 

\m

\n{\bf Remark 6} The previous two graphs, despite having coincident numbers of both vertices and edges, are not isomorphic even at the level of unlabelled graphs. 
This is clear from comparing valencies: Fig \r{S(4, 1)-Top-beta}.3)'s maximal vertex valency is 6, whereas \r{S(4, 1)-Top-beta}.4)'s single T vertex is of valency 7.   
%
{            \begin{figure}[!ht]
\centering
\includegraphics[width=1.0\textwidth]{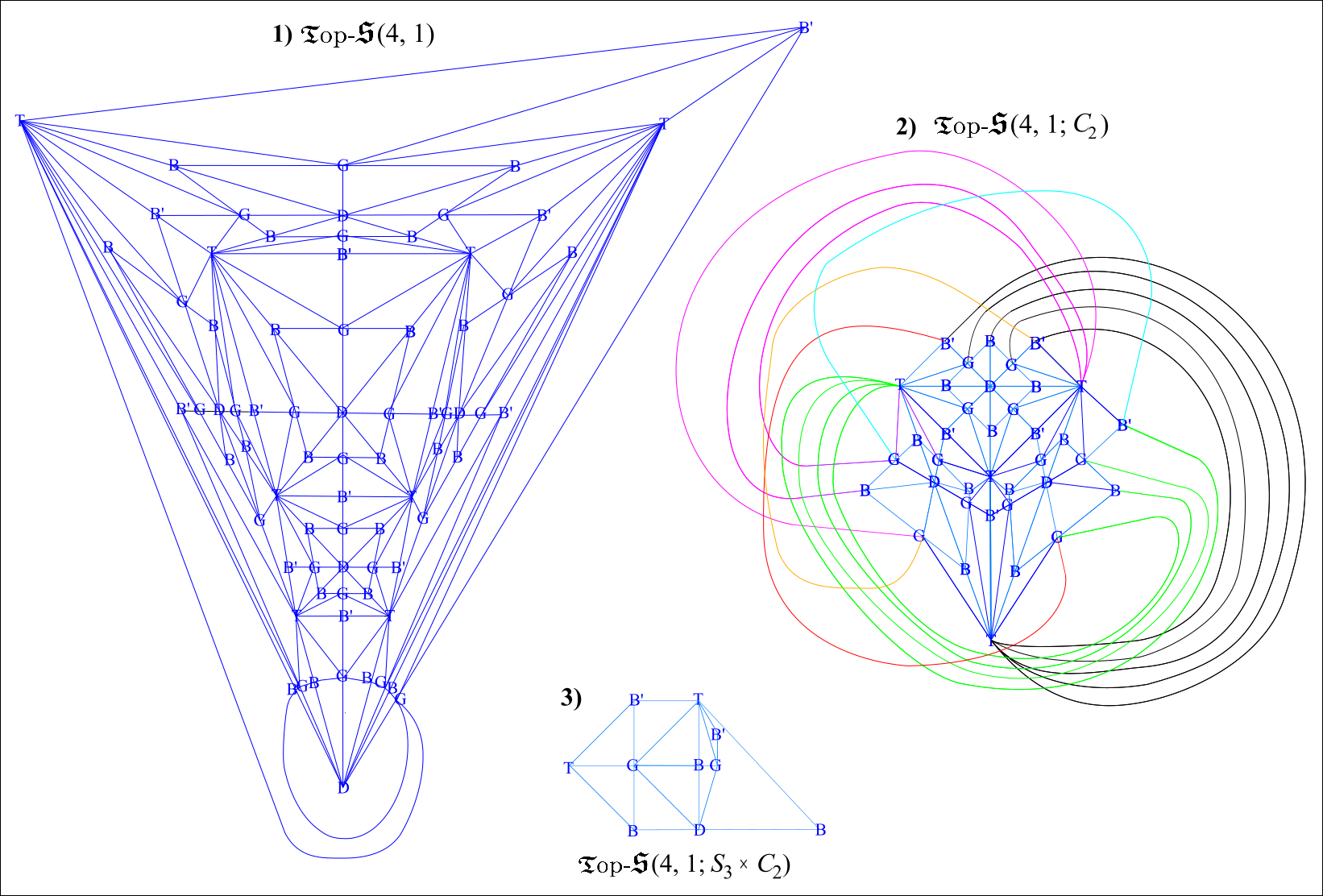}
\caption[Text der im Bilderverzeichnis auftaucht]{        \footnotesize{1) Planar graph representation of the 74-vertex cubic graph.  

\m 

\n 2) Graph representation of the 37-vertex $\mathbb{RP^2}$-embedded half-cube graph.

\m 

\n 3) Rectilinear representation of Fig 12.3)'s graph.   } }
\label{(4, 1)-Graphs} \end{figure}          }

\m 

\n{\bf Remark 7} By Remarks 1 and 2 of the previous section, for (3, 1) only $\FrS(3, 1)$ and $\FrS(3, 1; S_3)$ are finer than partitions, 
whereas for (4, 1) all shape space graphs are.
This is because all of the latter must possess at least 1 copy of each of $\mB$ and $\mB^{\prime}$.  
In particular, $\Leib_{\sFrS}(4, 1)$ is the first Leibniz space whose vertices are not just equivalent to $N$'s partitions (with the maximal coincidence-or-collision O removed).  
This is significant since spaces of partitions constitute a simpler and more commonplace object of study.
So, while study of the smallest few topological shape spaces partly reduces to the study of partitions, 
it is significant to note that this convenient reduction to a more established mathematical problem ceases to suffice from $N = 4$ upward.  

\m 

\n{\bf Remark 8} The (4, 1) shape-and-scale spaces' cones over all the previous largely do not produce any particularly simple, known or named graphs beyond 
noting these to be the cones over the previous. 
See moreover Fig \r{Complements} for various further complementary graph charaterizations of some of this section's shape(-and-scale) graphs.  
%
{            \begin{figure}[!ht]
\centering
\includegraphics[width=0.7\textwidth]{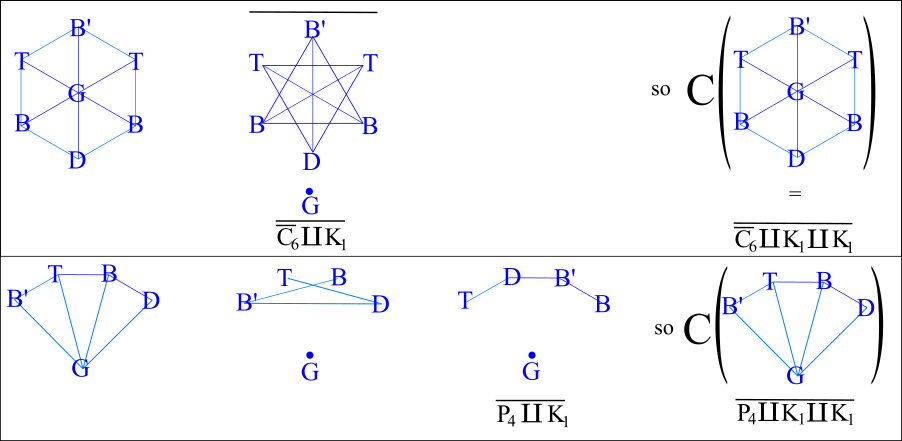}
\caption[Text der im Bilderverzeichnis auftaucht]{        \footnotesize{Some further complementary graph characterizations of (4, 1) shape(-and-scale) graphs.} }
\l{Complements} \end{figure}          }

\vspace{10in}

\section{(3, 2) Examples}\l{(3,2)}

{\bf Proposition 1} There are three topological types of 3-point configuration as per Fig \r{(3, 2)-Top-Shapes}.
These coincide with the 1-$d$ case's 3 topological classes, albeit with the G class now more broadly interpreted.  
%
{            \begin{figure}[!ht]
\centering
\includegraphics[width=0.40\textwidth]{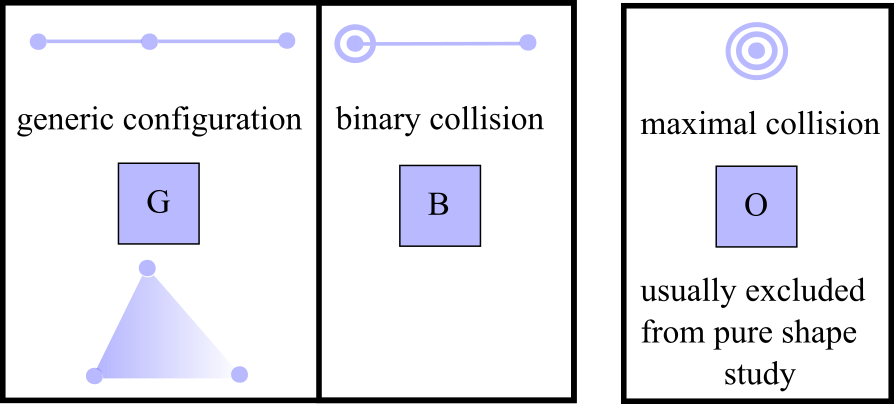}
\caption[Text der im Bilderverzeichnis auftaucht]{        \footnotesize{Topological classes of configurations for 3 particles in 1-$d$. 
Namely, the coincidence-or-collision-less generic configuration G, now covering both triangular and collinear metrically distinguished subcases, 
the binary coincidence-or-collisions B, and 
the maximal coincidence-or-collision O usually excluded from pure shape study.} }
\l{(3, 2)-Top-Shapes} \end{figure}          }

\m

\n{\bf Remark 1} As for (3, 1), these are all partitions, unless labels or mirror image distinctions further discern between types of B.  

\m

\n{\bf Remark 2} Regardless of whether the points-or-particles are labelled or mirror image configurations are held to be distinct, 
\be
\mbox{\#(G)} = 1 \m : 
\ee
the generic rubber triangle (including collinear cases but excluding binary or maximal coincidences-or-collisions) can be deformed from any labelling to any other.   
Thus topologically there is only one face or 2-cell.  
For a labelled triangle, regardless of whether mirror images are identified, 
\be 
\mbox{\#(B)} = \mbox{(ways of leaving one particle out)} 
             = 3                                            \m .  
\ee
The above represent two salient differences with the 1-$d$ case.  

\m 
  
\n{\bf Proposition 1} For distinguishably labelled points-or-particles, the maximal topological shape space is
\be
\Top\mbox{-}\FrS(3, 2)  :=  \Top\mbox{-}\FrS(3, 2; id) 
                         =  \mbox{claw}                 \m :
\ee
the 4-vertex claw graph alias 3-star $\mS_3$ graph with equally labelled `talons' as per Fig \r{S(3, 2)-Top}.1).     

\m

\Proof Continuity considerations show that these shapes fit together to form Fig \r{S(3, 2)-Top}.1).  $\Box$

\m

\n{\bf Proposition 2} For $(N, \, d) = (3, 2)$, the lattice of subgroup actions is as per Fig \r{(3, 2)-Latt}.  
%
{            \begin{figure}[!ht]
\centering
\includegraphics[width=0.05\textwidth]{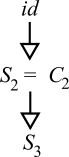}
\caption[Text der im Bilderverzeichnis auftaucht]{        \footnotesize{Lattice of distinct subgroup actions on $\sFrS(3, 2)$.} }
\l{(3, 2)-Latt} \end{figure}          }

\n{\bf Proposition 3} The other cases of shape space are as follows.  

\m

\n For precisely two indistinguishable points-or-particles, 
\be
\Top\mbox{-}\FrS(3, 2; C_2)  \es  \frac{\Top\mbox{-}\FrS(3, 2)}{C_2} 
                                \es  \m \mP_3                              \m : 
\ee
the 3-vertex path graph labelled as per Fig \r{S(3, 2)-Top}.2).

\m

\n For indistinguishable points-or-particles,   
\be
\Top\mbox{-}\Leib_{\sFrS}(3, 2)  :=  \Top\mbox{-}\FrI\FrS(3, 2; S_3)      
                                \es  \frac{\Top\mbox{-}\FrS(3, 2)}{S_3} 
                                \es  \mP_{2}                               \m : 
\l{Leib(3, 2)}
\ee
the 2-vertex path graph with distinct end-point labels as per Fig \r{S(3, 2)-Top}.3).
%
{            \begin{figure}[!ht]
\centering
\includegraphics[width=0.24\textwidth]{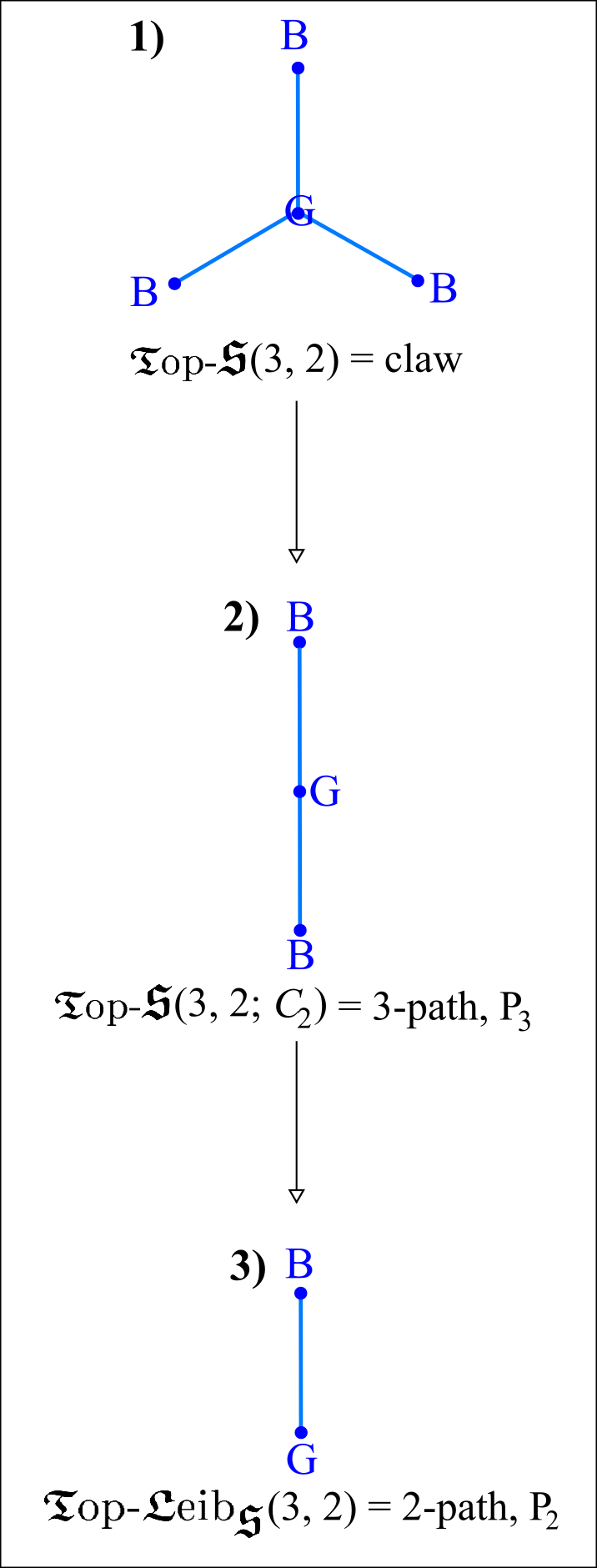}
\caption[Text der im Bilderverzeichnis auftaucht]{        \footnotesize{Corresponding lattice of (3, 2) topological shape space graphs.} }
\l{S(3, 2)-Top} \end{figure}          }

\m

\n{\bf Remark 3} When acting on the topological shape space claw graph, identifying mirror images has no separate effect.   
Thus 
\be
S_3 \times C_2 \m \mbox{ acts as } \m  D_3 \mbox{ (dihaedral group of order 6) } .  
\ee
This permits us to rewrite the labelling and mirror image based definition (\r{Leib(3, 2)}) of the topological configuration in space as 
\be
\Top\mbox{-}\Leib_{\sFrS}(3, 2)  \es  \frac{\Top\mbox{-}\FrS(3, 2)}{D_3}                                   \m . 
\ee
This is rather natural in configuration space since the unlabelled claw graph has automorphism group $D_3$. 

\m 

\n{\bf Remark 4} The above three graphs can also be characterized as $\mG$-apex cones over the discrete graphs $\mD_3$, $\mD_2$ and $\mD_1$, all labelled entirely with B's.  

\m

\n{\bf Remark 5} Among the normalizable shapes, this is just 3 points, corresponding to the 3 different labellings of the binary coincidence-or-collision, 
or just one point in the case in which these labellings are indistinguishable (Fig \r{R(3, 2)-Top}).  

\m

\n{\bf Proposition 4} For distinguishably labelled points-or-particles, the maximal topological shape-and-scale space is 
\be
\Top\mbox{-}\FrS(3, 2)  :=  \Top\mbox{-}\FrR(3, 2; id) 
                         =  \mbox{C(claw)}                 \m :
\ee
the cone over the claw graph as per Fig \r{R(3, 2)-Top}.1).     
%
{            \begin{figure}[!ht]
\centering
\includegraphics[width=0.45\textwidth]{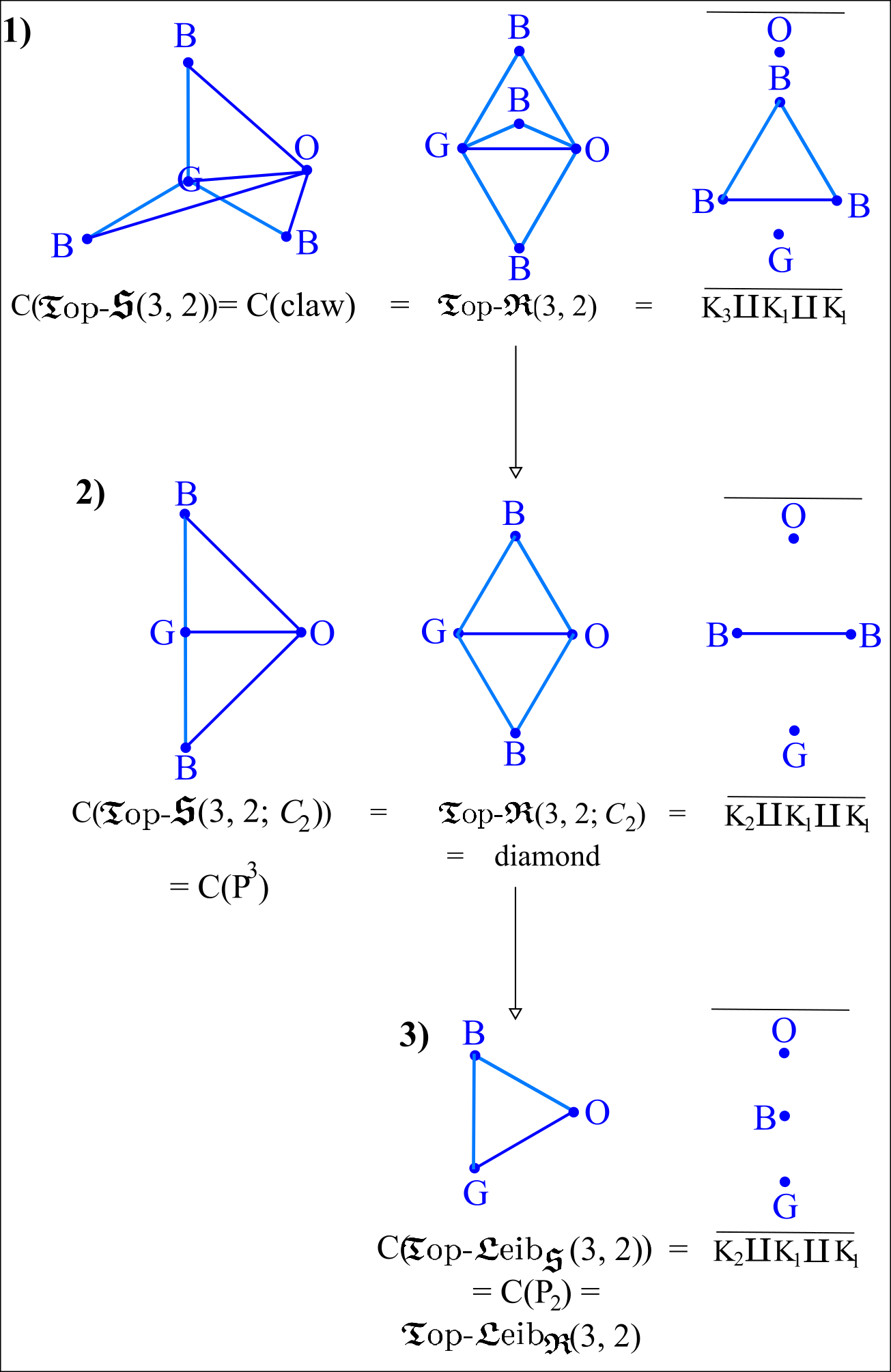}
\caption[Text der im Bilderverzeichnis auftaucht]{        \footnotesize{Corresponding lattice of (3, 2) topological shape-and-scale space graphs.} }
\l{R(3, 2)-Top} \end{figure}          }

\m 
 
\n{\bf Proposition 5} The other cases of shape-and-scale space are then as follows.  

\m

\n For precisely two indistinguishable points-or-particles, 
\be
\Top\mbox{-}\FrR(3, 2; C_2)  \es  \frac{\Top\mbox{-}\FrS(3, 2)}{C_2} 
                                \es  \m \mbox{ 2-fan } \mF_2 
= \mbox{ diamond }								\m : 
\ee
labelled as per Fig \r{R(3, 2)-Top}.2).

\m

\n For indistinguishable points-or-particles,   
\be
\Top\mbox{-}\Leib_{\sFrR}(3, 2)  :=  \frac{\Top\mbox{-}\FrR(3, 2)}{S_3} 
                                \es  \mC_{3}                               \m : 
\l{Leib-R(3, 2)}
\ee
the 3-cycle graph labelled as per Fig \r{R(3, 2)-Top}.3).

\m

\n{\bf Remark 6} In this case, note that all three shape-and-scale space graphs are complete tripartite.
This results in them being much more systematic and memorable if described in terms of their three-component complements.
In each case, two of these components are trivial $K_1$'s corresponding to O and G respectively, 
whereas the third component is the complete space with as many B vertices as the graph possesses ($K_1 = P_1$, $K_2 = P_2$, $K_3 = C_3$).
This reflects the cone-of-a-cone-graph Corollary of Sec 3.  

\vspace{10in}

\section{(4, 2) Examples}\l{(4,2)}

{\bf Proposition 1} For (4, 2), there are five topological types of configuration, as per Fig \ref{(4, 2)-Top-Shapes}.
%
{            \begin{figure}[!ht]
\centering
\includegraphics[width=0.6\textwidth]{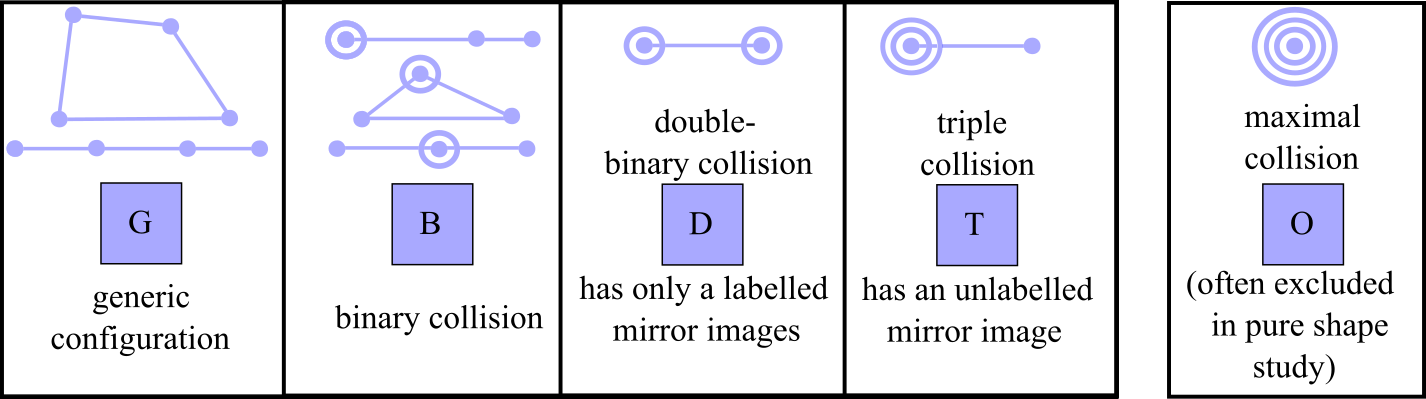}
\caption[Text der im Bilderverzeichnis auftaucht]{        \footnotesize{Topological classes of configurations for 4 points in 2-$d$: 
the maximal coincidence-or-collision O usually excluded from pure shape study, 
the ternary coincidences-or-collisions T, 
the double-binary coincidence-or-collision B, 
the binary coincidence-or-collision B,
and the coincidence-or-collision-less generic configuration G. } }
\label{(4, 2)-Top-Shapes} \end{figure}          }

\m

\n{\bf Remark 1} 
These coincide with (4, 1)'s 6 topological classes upon agglomerating B and $\mB^{\prime}$ since these now can be deformed into each other, 
and with the B and G classes elsewise more extensively interpreted.  

\m

\n {\bf Proposition 2)} These five topological types of configuration are in 1 : 1 correspondence with unordered partitions. 
Such a 1 : 1 correspondence holds moreover for any $N$ for $d \geq 2$. 

\m

\n{\bf Remark 2} I.e.\ G is 1 | 1 | 1 | 1, B is 2 | 1 | 1, D is 2 | 2, T is 3 | 1 and Q = O is 4. 

\m

\n {\bf Remark 3} Such a 1 : 1 correspondence does not hold for $d = 1$, $N \geq 4$ as Corollary 1 of the following Lemma.

\m 

\n{\bf Lemma 1} 
\be
\mbox{In 1-$d$ space, removing a point alters the topology of space}  \m .
\label{1-d-disconnects}
\ee
\n{\bf Remark 4} For the current article's 1-$d$ connected manifolds, removing a point from $\mathbb{R}^1$ disconnects it into two copies of $\mathbb{R}_+$, 
whereas it cuts $\mathbb{S}^1$ into a finite open interval.  

\m

\n{\bf Corollary 1} This causes multiplicity of i) mirror image distinct G shapes for $d = 1$ and $N \geq 2$

\m 

\n ii) Of mirror-image-identified G shapes for $d = 1$ and $N \geq 3$.  

\m 

\n{\bf Example 1} The ABC generic state is distinct from the ACB generic state by the binary collision in which B and C coincided to the left of A.  

\m 

\n{\bf Corollary 2} For $N \geq 4$, this results in collisions whose orders of occurrence along the topological line are topologically distinct, 
starting with the B to $\mB^{\prime}$ distinction for $N = 4$.  

\m

\n {\bf Remark 5} Regardless of whether the particles are labelled or mirror images are held to be distinct, 
\be
\mbox{\#(G)} = 1 \m : 
\ee
the rubber quadrilateral can be deformed from all labellings to all other labellings. 
Thus topologically there is only one 4-cell.  
This result holds moreover for $d \geq 2$ and any $N$.  
For $d = 1$, the larger multiplicity of cells with the top configuration space dimension is also rooted in (\ref{1-d-disconnects}).  

\m 

\n Also for a labelled quadrilateral, regardless of whether mirror images are identified, 
\be 
\#(\mB) = \left( C(4, 2) \mbox{ choices of pair } \right) 
        = 6                                                                      \m .  
\ee
\be 
\#(\mD) = \left( C(4, 2) \mbox{ choices of pair } \right)/2 \mbox{ orders } = 3  \m .  
\ee
\be 
\#(\mT) = \left( \mbox{ 4 ways of leaving 1 particle out } \right)  = 4          \m .  
\ee
\n{\bf Proposition 3} For distinguishably labelled points, the topological shape space is
\be
\Top\mbox{-}\FrS(4, 2) = \mC(\mbox{13-vertex tripartite graph})                  \m :
\ee
as exhibited and labelled in Fig \ref{S(4, 2)-Top}.1).     

\m 

\n{\bf Proposition 4} For $(N, \, d) = (4, 2)$, the lattice of subgroup actions is as per Fig \r{(4, 2)-Latt}.  
%
{            \begin{figure}[!ht]
\centering
\includegraphics[width=0.2\textwidth]{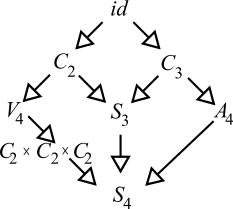}
\caption[Text der im Bilderverzeichnis auftaucht]{        \footnotesize{Lattice of distinct subgroup actions on $\sFrS(4, 2)$.} }
\l{(4, 2)-Latt} \end{figure}          }

\m 
  
\n{\bf Proposition 5} The other cases of (4, 2) topological shape space are as follows.  

\m 

\be
\Top\mbox{-}\FrS(4, 2; C_3) \m := \m \frac{\Top\mbox{-}\FrS(4, 2)}{C_3} \m = \m  \mF_4          \m : 
\ee
the 4-fan graph labelled as per Fig \ref{S(4, 2)-Top}.3), 
\be
\Top\mbox{-}\FrS(4, 2; A_4) \m := \m \frac{\Top\mbox{-}\FrS(4, 2)}{A_4} \m = \m \mW_4           \m : 
\ee
the 4-wheel graph labelled as per Fig \ref{S(4, 2)-Top}.6), 

\m 

\be
\Top\mbox{-}\FrS(4, 2; A_4) \m := \m \frac{\Top\mbox{-}\FrS(4, 2)}{A_4} \m = \m \mF_4           \m : 
\ee
another labelling of the 4-fan graph as per Fig \ref{S(4, 2)-Top}.7).  

\m 

\be
\Top\mbox{-}\Leib_{\sFrS}(4, 2)    :=     \Top\mbox{-}\FrS(4, 2; S_4) 
                                \m := \m  \frac{\Top\mbox{-}\FrS(4, 2)}{S_4} = \mbox{diamond}   \m : 
\ee
labelled as per Fig \ref{S(4, 2)-Top}.8), and three other cases as per Fig \ref{S(4, 2)-Top}.2), 4) and 5).  
%
{            \begin{figure}[!ht]
\centering
\includegraphics[width=0.7\textwidth]{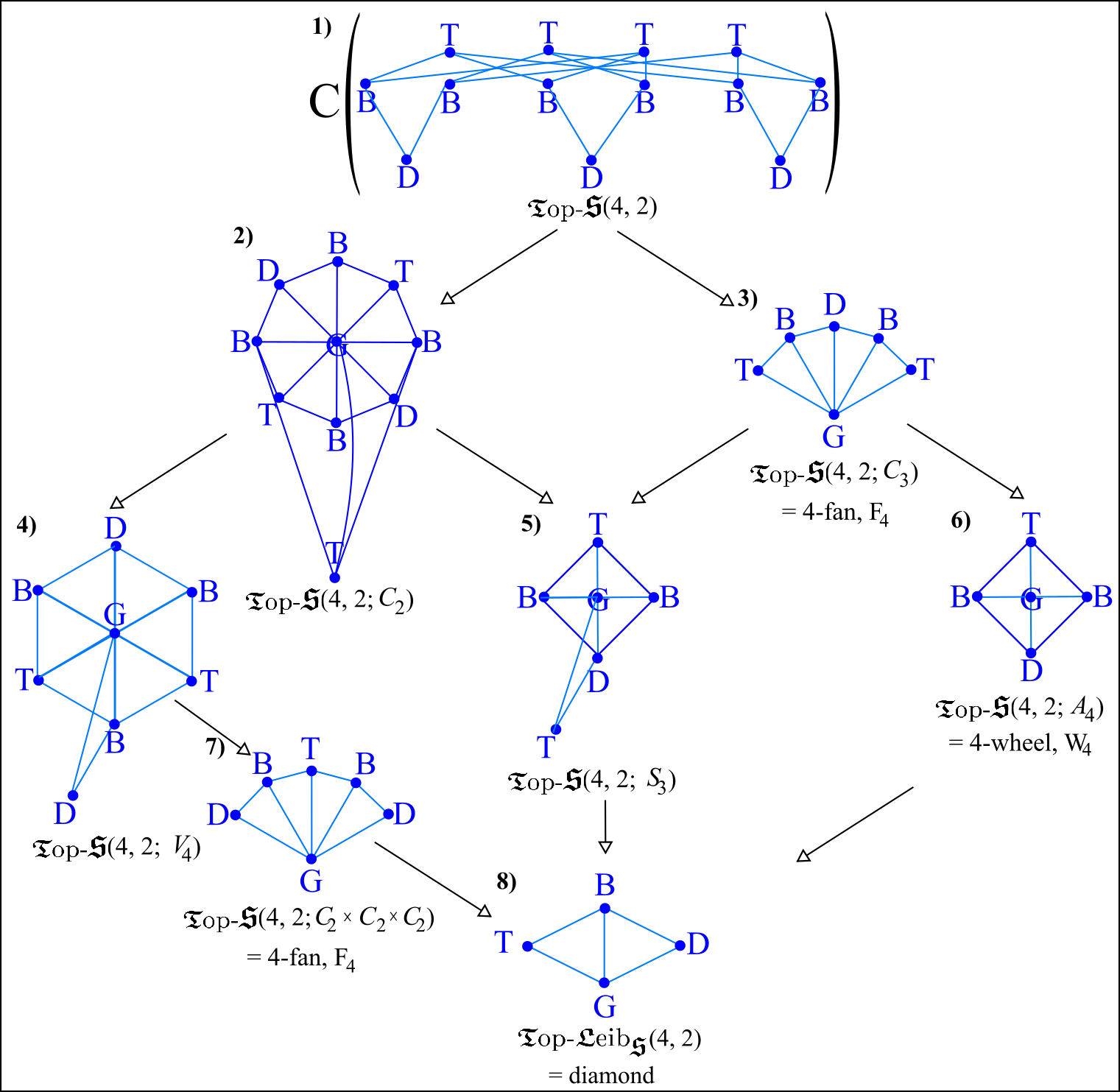}
\caption[Text der im Bilderverzeichnis auftaucht]{        \footnotesize{(4, 2) topological shape spaces for the lattice of differently-acting $\Gamma$ subgroups of $S_4$.} }
\l{S(4, 2)-Top} \end{figure}          }
%
{            \begin{figure}[!ht]
\centering
\includegraphics[width=0.52\textwidth]{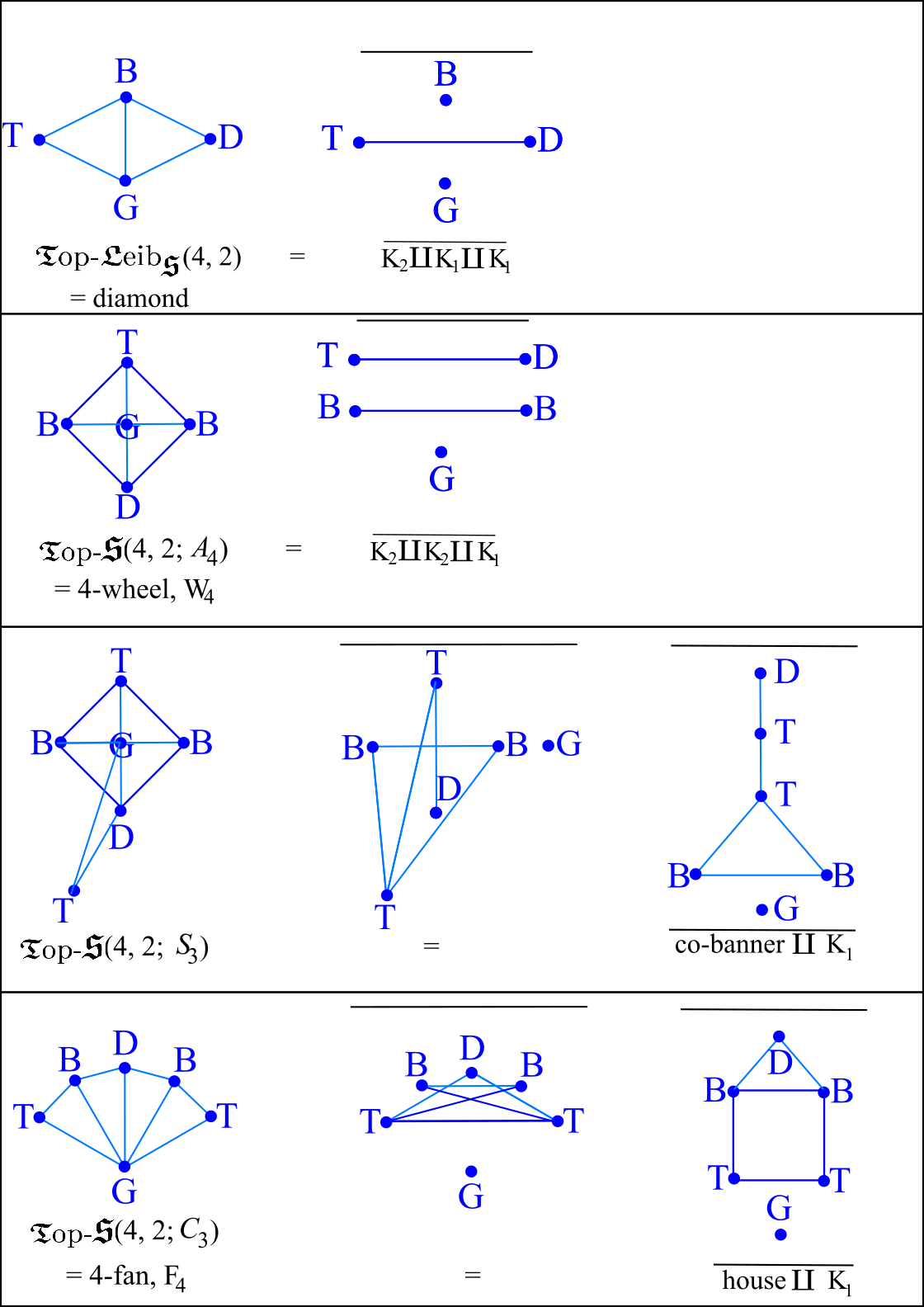}
\caption[Text der im Bilderverzeichnis auftaucht]{        \footnotesize{Complement graph presentation of various (4, 2) shape-and-scale spaces.} }
\l{R(4, 2)-Top} \end{figure}          }
 
\m

\n{\bf Proposition 6} For $N$ points in $d \geq 2$, the {\sl set} of topological configurations (including O) 
is in 1 : 1 correspondence with the set of all partitions of the corresponding $N$.  
The {\sl space} of topological shapes, however, has an additional structure -- the topological adjacency relation representable by graph edges -- 
which is not a structure that is usually ascribed to partitions.

\m 

\n{\bf Proposition 7} The corresponding scale-and shape relational spaces are the cones over Fig \ref{S(4, 2)-Top}'s graphs, with O as the new cone point. 
Cones over these mostly do not produce further distinctive notions or names of graphs.    
Four of the shape spaces admitting straightforward complement descriptions; 
the corresponding shape-and-scale spaces then admit complement descriptions consisting of each of these alongside an O $\mK_1$ singleton, as per Lemma 2 of Sec 3.  

\m 

\n{\bf Corollary 3} For $d \geq 2$, i) 
\be
|\Top\mbox{-}\Leib_{\sFrR}(N, \, d)| = p(N) \mbox{ } : 
\ee  
the number of partitions of $N$ into natural numbers.  

\m 

\n ii) 
\be
|\Top\mbox{-}\Leib_{\sFrS}(N, \, d)| = p(N) - 1 \mbox{ } : 
\ee
\n{\u{Proof}} i) follows from Lemma 1.  ii) then follows by exclusion of the maximal coincidence-or-collision O.  $\Box$ 

\m

\n {\bf End-Note 1} While one can still use P to denote pentuple collision, H to denote hexuple collision...  
the unordered partitions themselves provide a more satisfactory and ultimately necessary notation for collisions in $d \geq 2$. 
1-$d$ is then furthermore permissive of ordered partition distinctions, whether or not with mirror image identification.

\section{Topological shape spaces within metric shape spaces}

\n{\bf Example 1} The $\Top\mbox{-}\FrS(3, 2)$ claw graph and the cone thereover becomes each of the following. 
In each case, `equal masses' are assumed.

\m

\n a) In the $\mathbb{R}^2$ pure-shape mirror-images-distinct case, its central G vertex becomes $\mathbb{S}^2$, 
minus the three talon B-points that are evenly spaced out along its equator; 
see Fig \ref{Last-Up}.a) and Fig 1 for further details.

\m

\n b) In the $\mathbb{R}^2$ pure-shape mirror-images-identified case, its central G vertex becomes the hemisphere with edge included, $\mathbb{S}^2_0$, 
minus the three talon B-points that are evenly spaced out along its bounding equator; 
see Fig \ref{Last-Up}.b).  

\m 

\n c) In the $\mathbb{R}^2$ scaled mirror-images-distinct case, its central G vertex becomes $\mathbb{R}^3$, 
minus the three talon B-lines emanating at $2\pi/3$ to each other in the equatorial plane, 
and the O point at which the three join.
This O point moreover has the status of a separate stratum. 
Thus in this case the stratum-by-stratum split is a coarse-graining of the topological shape graph into apex vertex O on the one hand, 
and non-apex vertices G and B treated together on the other.  
See Fig \ref{Last-Up}.c).

\m 

\n d) In the $\mathbb{R}^2$ scaled mirror-images-identified case, its central G vertex becomes the half-space $\mathbb{R}^3_0$, 
minus the three talon B-lines emanating at $2\pi/3$ to each other in the bounding equatorial plane, 
and the O point at which the three join.
This O point again has the status of a separate stratum, and the preceding coarse-graining comment applies again. 
See Fig \ref{Last-Up}.d).

\m

\n e) In $\mathbb{R}^3$, mirror-images-identification is obligatory.  
The pure-shape case's central G vertex becomes the open hemisphere $\mathbb{S}^2_+$ and all of a separate $\mathbb{S}^1$ stratum bar three equally spaced out points.
The three talon B-points constitute these remaining three points. 
See Fig \ref{Last-Up}.e).

\m 

\n f) In the scaled $\mathbb{R}^3$ case, the central G vertex becomes the open half-space $\mathbb{R}^3_+$ 
and all of a separate $\mathbb{R}^2$ stratum bar the three talon B-lines emanating at $2\pi/3$ to each other 
and the O point at which the three join.
The three B-lines form part of a punctured plane $\mathbb{R}^2_*$ stratum, 
whereas the puncture O itself is a separate stratum.  
See Fig \ref{Last-Up}.f).

\m 

\n{\bf Remark 1} Thus in cases e) and f) the topological graph split is {\sl not} aligned with the stratum-by-stratum split.  

\m 

\n{\bf Remark 2} Stratification is here due to the full $SO(3)$ only acting on non-collinear configurations, 
with just an $SO(2)$ subgroup acting on collinear non-maximal coincidence-or-collision configurations, 
and merely an $id$ subgroup acting on the maximal coincidence-or-collision configuration O.  
In \cite{A-Monopoles} this was moreover characterized this as an example of {\it trivially-contiguous} stratification:  
manifolds (or orbifolds) with boundaries (and corners etc) in which some of the boundaries etc are geometrically distinct, 
and yet remain contiguous to the top stratum in the manner of manifold geometry.

\m 

\n g) For 3 points in $\mathbb{T}^2$ -- the toroidal triangles \cite{ATorus} -- including scale is obligatory. 
In this case, there is only one stratum, which is topologically and flat-metrically $\mathbb{T}^4$.  
This generically consists of G configurations, with 3 $\mathbb{T}^2$'s therein corresponding to the B configurations, 
triple-touching at a single point corresponding to the O configuration
See Fig \ref{Last-Up}.g).

\m

\n h) For 3 points in $\mathbb{S}^2$ -- the spherical triangles \c{Kendall87, Kendall, FileR, ASphe} -- including scale is once again obligatory. 
G corresponds to the principal stratum, which is topologically and metrically (a compact model for) $\mathbb{H}^3$ \cite{Kendall}. 
Four discrete points do not pertain to the principal stratum (which is thus, more accurately, $\mathbb{H}^3\ \mbox{4 points}$): O and the 3 antipodal realizations of the B's.  
The rest of the B's form circles minus two points (a shared O and each's particular antipodal B), pertaining entirely to the principal stratum. 
Thus in this case, the stratification fine-grains a topological class -- B -- into subclasses realized respectively within distinct strata.  
See Fig \ref{Last-Up}.h).

\m 

\n{\bf Remark 3} That topological classes need not coincide with stratificational classes can be accounted for by group actions having a say in the latter.  
%
{            \begin{figure}[!ht]
\centering
\includegraphics[width=1.0\textwidth]{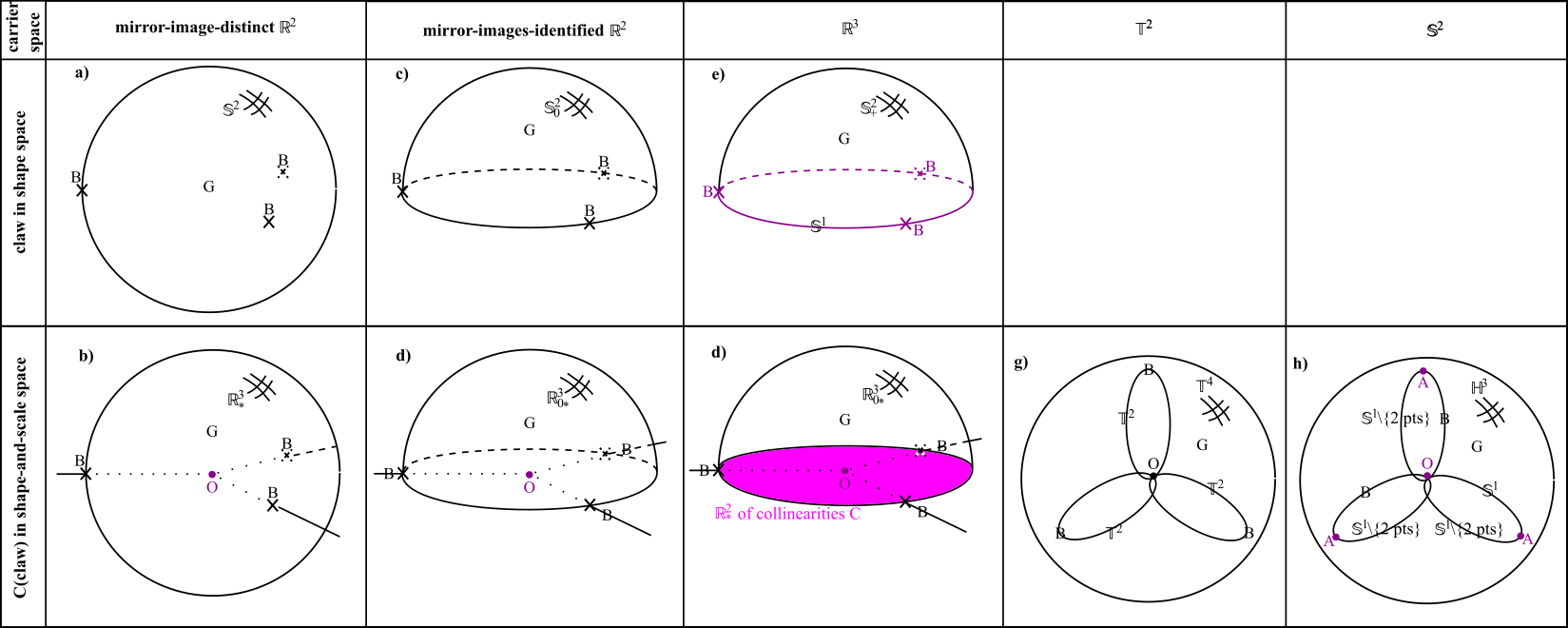}
\caption[Text der im Bilderverzeichnis auftaucht]{        \footnotesize{Eight realizations of claw or the cone thereover.
Separate strata are indicated in indigo (if bottom strata) or in magenta (if intermediate strata). 
%
}  }
\l{Last-Up} \end{figure}          }

\m

\n{\bf Example 2} Fig \r{S(4, 1)-Met-Top} shows how (4, 1)'s 74-vertex regular refinement of the cube is also realized as a tessellation of the metric-level shape space manifold.
%
{            \begin{figure}[!ht]
\centering
\includegraphics[width=0.9\textwidth]{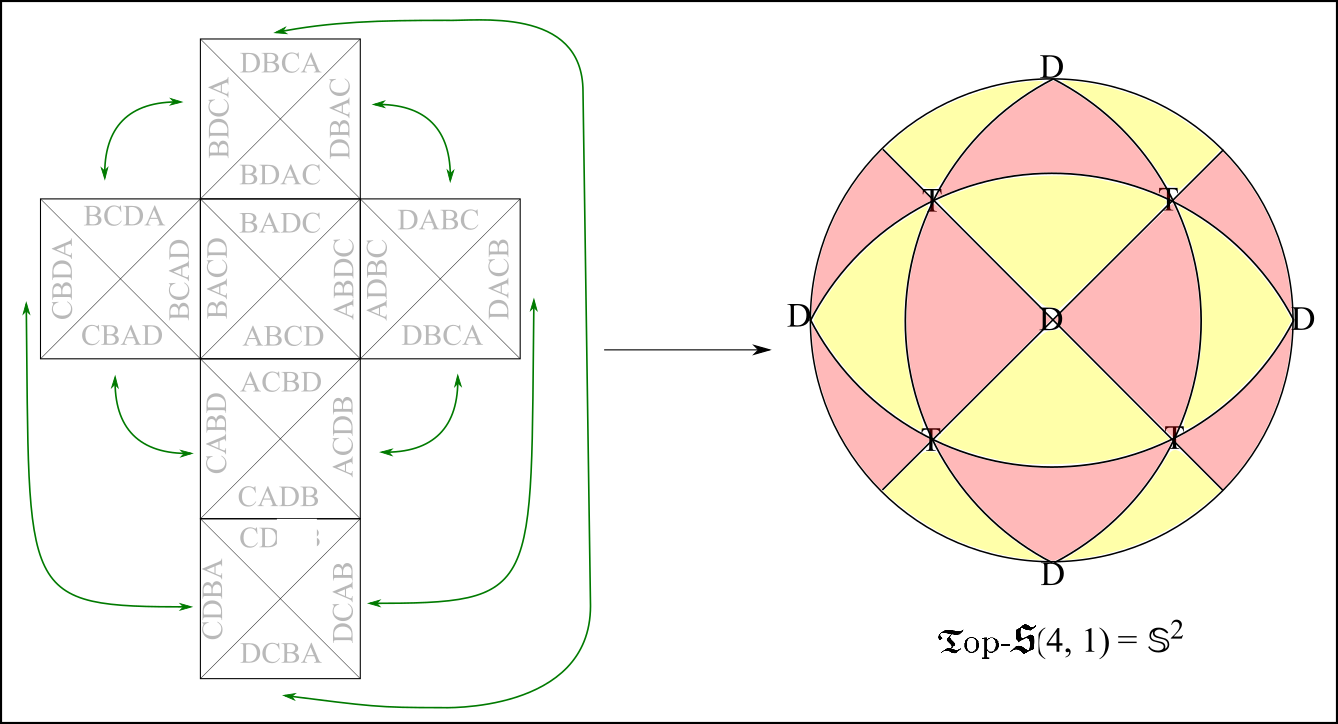}
\caption[Text der im Bilderverzeichnis auftaucht]{        \footnotesize{(4, 1) topological-level structure of the metric-level shape space manifold.  
%
}  }
\label{S(4, 1)-Met-Top} \end{figure}          }

\m 

\n{\bf Example 3} See \cite{IV} for the metric shape space realization of the (4, 2) topological shape space and \cite{Affine-Shape-2} for its affine shape space re-realization. 

\vspace{10in}

\section{Probability and Statistics on topological shape spaces}

\n{\bf Structure 1} One natural measure at the level of finite graphs is the {\it uniform measure}, in the sense of 
\be 
\mbox{Prob(vertex V)} = \frac{1}{|\mG|} \m \m \forall \, \, \mV \, \in \, \mG
\ee 
for $|\mG|$ the number of vertices in the graph $\mG$. 
If one uses this for topological shape graphs, 
it is considerably simpler as regards setting up the corresponding Probability and Statistics theory, and computing out examples, than its geometrical shapes counterpart.  

\m

\n{\bf Remark 1} On the one hand, this has some pedagogical value: measure theory, probability and statistics on differentiable manifolds (let alone stratified manifolds) 
is {\sl well beyond} the grasp of undergraduates, and indeed of most graduates that do not specifically specialize in Shape Statistics (or stratified manifolds, or 
Statistics--Topology interplay \cite{ST1, ST2})

\m 

\n{\bf Remark 2} On the other hand, presents a `modelling discontinuity' upon passing to the metric version, 
since $G$ is of generic dimension whereas all collisions are of a smaller non-generic dimension. 
(This holds for all Similarity Shape Theories and Euclidean Shape-and-Scale-Theories, but not necessarily for all strata in their affine counterparts: see \cite{Affine-Shape-1}.)

\m

\n{\bf Remark 3} A further compromise in introductory-level  pedagogy is then to also use the geometrically-standard measure on $\mathbb{S}^2$ in the context of the shape sphere 
as a separate source of examples.

\m

\n{\bf Remark 4} Returning to topological shape space models themselves, it is only slightly more technically involved to allot different weights to vertices.
Discrete probability distributions on the vertices of graphs are straightforward both to set up and to do calculations with.  

\m 

\n{\bf Structure 2}
\be 
\mbox{Prob(vertex V)} = w(\mV) 
\ee 
such that 
\be 
\sum_{\sV \, \in \, \sG} w(\mV) =  1  \m . 
\ee 
\n {\bf Structure 3} {\sl Standard} discrete Statistics \cite{CB-Book} is then suitable for the current Section's treatment.

\section{Dynamics and Quantum Theory of topological shapes(-and-scales)}

\subsection{Classical Markov chain modelling}

\n One straightforward approach to this is using Markov chains, for which nonzero probabilities are allotted in both directions along the edges of topological adjacency 
of the topological shape(-and-scale) space graphs.
These give evolution over a series of time-steps, with long-term outcomes of where the state of the system ends up computible.    

\m 

\n{\bf Structure 1} A simple and in some ways natural set-up for this is to assign uniform probabilities 
to all edges in a given topological shape(-and-scale) space graph (Fig \r{Markov}.1).

\m 

\n{\bf Structure 2} Another is to additionally incorporate staying probabilities (loops from each vertex to itself) to model remaining in that same state during 
the next `dynamical time step' (Fig \r{Markov}.2).
%
{            \begin{figure}[!ht]
\centering
\includegraphics[width=0.5\textwidth]{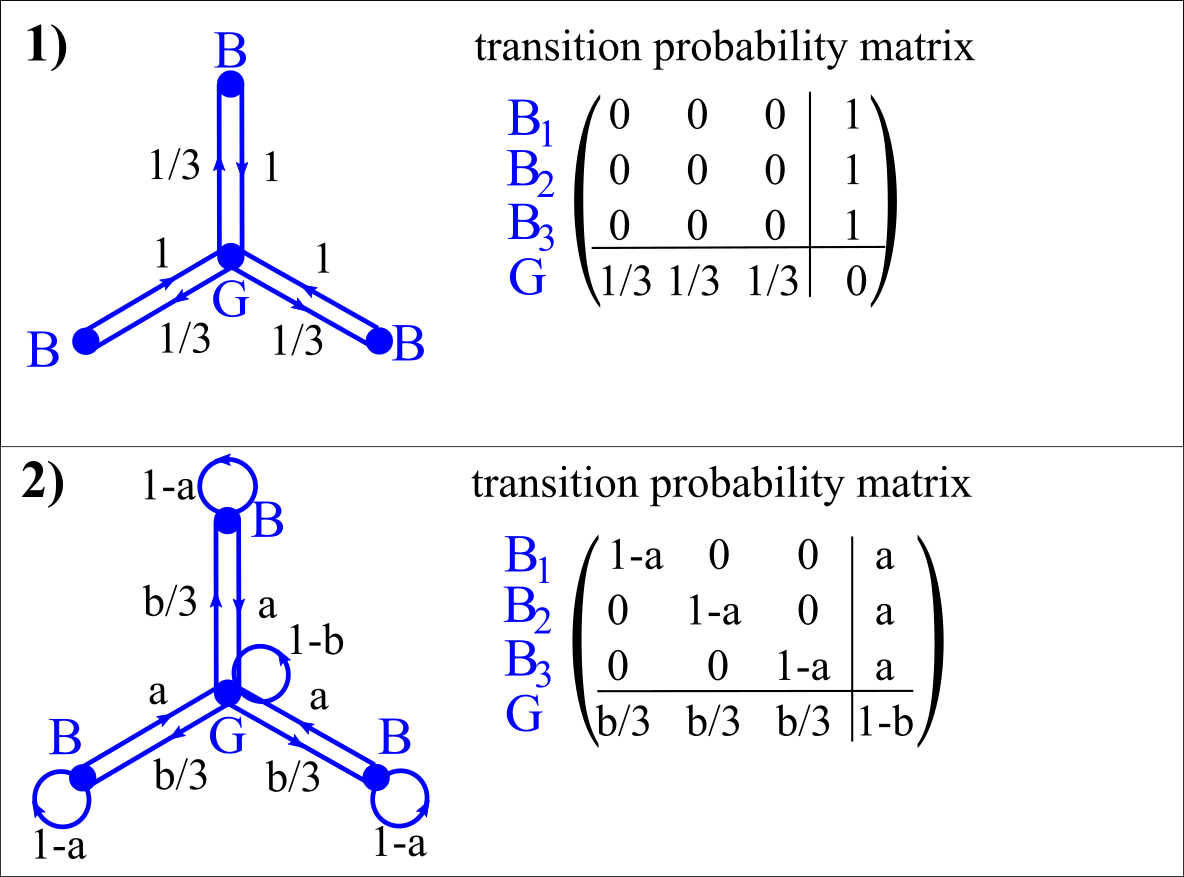}
\caption[Text der im Bilderverzeichnis auftaucht]{        \footnotesize{Markov chain set-up for $\Top\mbox{-}\FrS(3, 1) =$ claw, 1) without and 2) with staying probabilities.} }
\l{Markov} \end{figure}          }

\subsection{Classical Lagrangian modelling}

For the sake of familiarity, we preliminarily postulate a Lagrangian
\be
{\cal L}(\mG) \es \frac{1}{2}\sum_{v \in v(\sG), \, e \in e(\sG)} \frac{\Delta x_{ve}}{\Delta t} \frac{\Delta x_{ve}}{\Delta t} - V(x_v)          \m  
\label{Lag}
\ee
for our discrete graph-theoretic dynamics, which we restrict in the standard dynamical manner to being quadratic. 
$V = V(v_e \mbox{ alone})$ is here some potential function dependent on the vertices alone, i.e.\ the discrete graph-theoretic analogue of a velocity-independent potential.  
The {\it graphical velocities} we introduce here are to be {\it incidence matrices} 
\be 
\frac{\Delta x_{ve}}{\Delta t}  =  M_{ve} 
                               :=  \left\{\s{  \s{  \mbox{ $-1$ if edge $e_{ij}$ leaves vertex $v_i$ }  }
							                     {  \mbox{ $1$  if edge $e_{ij}$ enters vertex $v_i$ }  }  }{  \mbox{0 otherwise}  }  \right.  \m .
\label{Mve}
\ee 
These confers directionality to one's edges; for an undirected version -- a symmetrical treatment of each -- the two signs are positive. 

\m 

\n In fact, we will not usually want an external, background or absolute time notion such as this $t$ in our theory. 
We get around this by postulating -- instead of a discrete graph-theoretic Lagrangian -- a {\it graphical Jacobi arc element}
\be
\Delta {\cal J} = 2 \sqrt{\frac{1}{2}\sum_{v \in v(\sG), e \in e(\sG)} {\Delta x_{ve}}{\Delta x_{ve}}} \, \sqrt{E - V(x_v)}                        \m , 
\ee 
which is built out of {\it timeless changes} $\Delta x_{ve}$ rather than velocities.  
Following \cite{TRiPoD}'s continuum counterpart, the corresponding formula for the {\it graphical momenta} is 
\be
p_{ve}  \:=  \frac{\Delta \, \Delta{\cal J}}{\Delta \, \Delta x_{ve}} 
        \es  \frac{\sqrt{E - V(x_v)} \Delta x_{ve}}{\sqrt{\frac{1}{2}\sum_{v \in v(\sG), \, e \in e(\sG)} {\Delta x_{ve}}{\Delta x_{ve}}}}
		\=:  \frac{\Delta x_{ve}}{\Delta t^{\se\sm}}
\ee 
for $t^{\se\sm}$ the {\it graphical emergent Machian time}.  
This replaces eq.\ (\ref{Lag})'s $t$. 
Via the emergent Machian version of the first equality of (\r{Mve}), we identify the classical graphical momenta themselves to be 
\be 
p_{ve} = M_{ve}                                                                                                                                    \m : 
\ee 
incidence matrix `arrows'.  

\m 

\n There are a couple of subtleties in the above formulation. 
The most basic Lagrangian from Optimization is (linear isotropic cost)
\be 
{\cal L} = \mbox{tr} \, A
\ee
for $A$ the {\it adjacency matrix} encoding the graph's edges 
\be
A_{ij}  :=  \left\{ \s{\mbox{1 if there is an edge from vertex $v_i$ to vertex $v_j$}}{\mbox{0 otherwise}} \right.  \m .  
\ee 
Edges are a priori most natural as change or momentum variables; 
however, linear Lagrangians of this kind do not permit a Jacobian formulation. 
One way out would be to evoke `quadratic isotropic cost', such as   
\be 
{\cal L} = \mbox{tr}(A^2) \m .  
\ee 
The choice we make, however, is motivated by two features of the {\it graphical Laplacian matrix} 
\be
\triangle(\mG)_{vv^{\prime}} = D_{vv^{\prime}} - A_{vv^{\prime}}                      \m , 
\ee 
where 
\be 
D_{ij}  :=  \mbox{diag}(d(v_i))                                                                                      \m , 
\ee 
is the {\it degree matrix}. 
Firstly, this is a corrected $A$ rather than a corrected $A^2$. 
Secondly, it is none the less expressible as the square of a further quantity,  
\be 
\triangle(\mG)_{vv^{\prime}} = M_{ve}^{\sT} \, M_{ev^{\prime}} \m ,  
\ee 
so we treat this `square root of a corrected adjacency matrix of edges' $M_{ve}$ as the basic change-and-momentum variable of our system, as postulated.

\m 

\n We end by giving the corresponding {\it discrete graph-theoretic Hamiltonian} is, via the {\it discrete graph-theoretic Legendre transformation}   
\be 
{\cal H}(\mG)  \:=  p_{ve}^{\sT} \, \frac{\Delta x_{ve}}{\Delta t^{\se\sm}} - {\cal L}(\mG) 
               \es  p_{ve}^{\sT} \, p_{ve} - \left( \frac{1}{2} \, p_{ve}^{\sT} \, p_{ve} - V(x_v) \right) 
			   \es  \frac{1}{2}  \, p_{ve}^{\sT} \, p_{ve} + V(x_v)  
               \es  \frac{1}{2}  \, M_{ve}^{\sT} \, M_{ve} + V(x_v) 			                                                                   \m .  
\ee 
We favour the current subsection's development over the previous as regards 
a) developing Background-Independent Physics -- including finding an emergent time therein rather than assuming a background time --
and b) proceeding onto a quantum treatment as follows.

\subsection{Quantum-mechanical modelling} 

We outline here an approach to Quantum Mechanics for fixed finite discrete graph. 
This is to be applied to each topological shape(-and-scale) graph in a second article \c{QM-Top-Shapes}.  
The state space are finite vectors whose components are the vertices of the graph. 
We need to promote the classical Hamiltonian to an operator acting on this. 
One natural way of assigning this is to untrace the classical free Hamiltonian to provide the graphical Laplacian matrix $\triangle(\mG)_{vv^{\prime}}$. 
We furthermore scale this with a measure of noncommutativity, 
\be 
\widehat{\cal H}^{\sf\sr\se\se}_{vv^{\prime}} = k^2 \triangle(\mG)_{vv^{\prime}}                                                              \m .
\ee 
It is entirely straightforward to carry over the potential term to the quantum realm as well, 
\be 
\widehat{\cal H}_{vv^{\prime}} = k^2 \triangle(\mG)_{vv^{\prime}} + \delta_{vv^{\prime}} \, V(x_{v})                                          \m .
\ee 
This gives a quantum equation 
\be 
\widehat{\cal H}_{vv^{\prime}} \Phi_{v^{\prime}} = k^2 \triangle(\mG)_{vv^{\prime}} \Phi_{v^{\prime}} + V(x_{v}) \Phi_{v} = E \, \Phi_{v}     \m . 
\ee 
This is of time-independent alias stationary Schr\"{o}dinger equation form; 
we solve these for energy eigenspectra and finite wavefunction vertex vectors in \c{QM-Top-Shapes}.  

\m

\n This is not to be confused with Pauling's QM on graph models of molecules \cite{Pauling} or similar work concerning networks of thin wires \cite{QM-Graphs}, 
as these are metric QM's. 
Nor is it to be confused with Freedman et al's QM of graphs \cite{QM-Graph-2}, for which the state space is a variety of graphs rather than a single fixed graph. 
However, once one begins to consider dynamical shape(-and-scale) theories, such as with variable particle numbers or changing automorphism groups, 
then features of Freedman's approach -- restricted to bona fide shape(-and-scale) graphs -- become applicable as well.

\subsection{Types of question that can be posed of each such model}

\n 1) {\bf Time step by time step questions}. 

\m 

\n Starting from initial state before the first time-step, what is 
\be 
\mbox{Prob($n$th timestep lands the system in state V)?}
\ee 
Or its generalization 
\be 
\mbox{Prob($n$th timestep lands the system in within subgraph G$^{\prime}$)?}
\ee 
\n 2) {\bf End-state questions}. 
E.g.\ what is 
\be
\mbox{Prob(system ends up in the generic state)?}
\ee 
\n 3) {\bf Timeless questions}. 
Topological Shape Theory can be viewed as a classical timeless records model, as regards investigation of questions such as what is 
\be 
\mbox{Prob(generic shape $G$)}  
\ee 
{\sl without any reference to when}.  

\m

\n As one motivation, cosmological analogues include what are 
\be 
\mbox{Prob(universe is flat)}  \m ,
\ee 
\be 
\mbox{Prob(universe is isotropic)}  \mma \mbox{ and }
\ee 
\be 
\mbox{Prob(universe is homogeneous) } ?   
\ee 
Aside from the classical and observational interest in such questions (and their quantification by small tolerance parameters), 
noted theoreticians such as Hawking, Page, Unruh and Wald \cite{H84, HP86, HP86, UW89} put forward a \NSI scheme for computing such probabilities at the level of Quantum Cosmology.  
Thus began the `timeless approaches' to the Problem of Time, so a second motivation for considering timeless questions comes from the Foundations of Quantum Gravity. 

\m

\n Similarity Shape Theory analogues of this include 
\be 
\mbox{Prob(triangle model universe is near-equilateral)}
\ee 
an opposite to which is Kendall's 
\be 
\mbox{Prob(triangle model is near-collinear)}
\ee 
in the context of sampling in threes upon location data to see if a statistically significant number of triples are approximately aligned. 
This was used for instance to assess the standing stones of Cornwall \cite{Kendall84, Kendall} and claims of quasar alignments \c{Kendall87, Kendall89}.   
Kendall's Shape Statistics provides a {\sl classical} computational scheme for such probabilities, 
which the Author identified \cite{AKendall, ABook} as of value as a {\it computationally explicit classical precursor} for the Foundations of Quantum Gravity and Quantum Cosmology. significant topic of 
Timeless Records Theory \c{GMH, H99, H03, Records, H09, AKendall, ABook}.  
The current article notes that while Kendall-type schemes take an increasingly nontrivial amount of mathematics to set up 
as whichever of $d$, $N$ and the complexity of the automorphism group increase, 
Topological Shape(-and-Scale) Theory provides {\sl very} mathematically simple working models of relationally or Background Independence significant 
models of classical and quantum Timeless Records Theory. 
These have a topological shapes analogue when involving purely topological properties, 
so a few such questions can be answered at both the geometrical and topological levels.

\m

\n We next turn to a further source of diversity in the timeless approaches literature: conditional probabilities questions: 
the Page--Wootters \cite{PW83}, Page \cite{Page1, Page2} and Gambini--Porto--Pullin \cite{GPP04, GPPT} approaches. 
These are once again QM-specific schemes.

\m

\n Cosmological analogues of this now include 
\be 
\mbox{Prob(universe is flat | it is isotropic)}  \m ,
\ee 
\n On the other hand, Similarity Shape Theory analogues of this include  
\be
\mbox{Prob(triangle is approximately isosceles | it is approximately collinear)}  \m ,
\ee 
for which \c{A-Pillow, Max-Angle-Flow} provide a trove of worked examples at the classical level.  

\m 

\n The current article's work leads to {\sl many} topological shape(-and-scale) questions of the present subsection's kind, 
which are moreover readily answerable using whichever of Classical Dynamics, Probability or stationary QM are appropriate.  

\vspace{10in}
 
\section{Conclusion}

Kendall-type Geometrical Shape(-and-Scale) Theories \cite{Kendall, FileR, Bhatta, PE16, ABook} are based on constellational primality 
and quotienting out geometrical automorphism groups.  
In the current article, we have presented a simpler notion of Shape(-and-Scale) Theory, of rubber shapes with coincidence primality 
(collision primality if the constellation's points are material particles).  
This gives Topological Shape(-and-Scale) Theory, 
one model for which can concurrently underlie many models of its Geometrical counterpart that differ as regards the further geometrical structure attributed. 
In this regime, a given Geometrical Shape(-and-Scale) Theory's continuous geometrical automorphism group vanishes from contention, as do almost all of the features 
of the carrier space (absolute space when physically implemented) that the constellations in question are given upon. 
The only features of connected manifold-without-boundary carrier spaces which survive are whether the dimension is $\geq 2$ or just 1, and, in the latter case, 
whether the model's carrier space is open -- $\mathbb{R}^1$ -- or closed: $\mathbb{S}^1$.  
We accounted for this distinction in basic topological terms. 
On the other hand, whether or not the model has scale survives as a feature, through its connection with whether or not the model's configurations 
can include the maximal coincidence-or-collision.
Finally, mirror image identification and particle label distinguishability features -- discrete automorphisms -- transcend to the topological shape(-and-scale) level.  

\m 

\n On the one hand, Geometrical Shape(-and-Scale) Theory's configuration spaces -- shape(-and-scale) spaces -- 
are in general \c{Kendall, GT09, PE16, KKH16} stratified manifolds \c{ABook, Pflaum, BanaglBook, Kreck}. 
These are objects largely beyond the scope of current mathematics and certainly beyond the conventional toolkit of theoretical physicists 
(see \cite{ABookApp} and the last Appendix in \c{A-Monopoles} for a brief introduction resting upon Theoretical/Mathematical Physics familar material).
This renders study of almost all Geometrical Shape(-and-Scale) Theories hard (including to physicists, which is of concern since, firstly these theories are excellent 
models of many aspects of Background Independence, and, secondly, stratification recurs \cite{DeWitt70, Fischer70, Fischer86, FM96, Giu06, ABook} 
in the study of GR's own reduced configuration spaces such as Wheeler's Superspace \cite{Battelle}).  
One partial way out is that a few shape spaces are just manifolds: the spheres and complex-projective spaces alluded to in Sec 2. 
On the other hand, Topological Shape(-and-Scale) Theory's configuration spaces are just graphs: {\sl very} mathematically accessible objects, 
moreover not requiring an undue amount of Graph Theory to understand (one of \c{I}'s Appendices will largely do for the current article's scope).  
The current article shows that this provides a formidable further diversity of examples, 
even just working up to particle number 4 and dimension 2. 

\m 

\n Topological Shape(-and-Scale) Theory's model examples are moreover of likely practical and pedagogical use in Shape Theory's hitherto largest application: Shape Statistics.
The latter is clear from our outline of the far greater ease with which Probability and Statistics can be set up on a graph than a manifold (let alone a stratified manifold).  
On the other hand, while doing Dynamics or QM on graphs is ab initio less familiar, we also outlined a straightforward manner in which this can be done 
(and \c{QM-Top-Shapes} is a sequel in this direction).

\m 

\n Let us end by commenting on the field of study called Shape(-and-Scale) Theory amounting to an exposition of models of Background Independence 
of relevance to deep physical themes such as the Absolute versus Relational motion debate and both the foundations and dynamics of each of General Relativity and Quantum Gravity.
We have documented how the relational side of the debate is realized by a portmanteau of shape(-and-scale) models.  
The current article's Topological Shape(-and-Scale) Theory is of particular interest in giving both a more solvable and less a priori structured paradigm of Background Independence.
Interesting past literature this complements includes the Topological Background Independence considerations of 
Isham of quantization at the topological-space and metric-space levels \cite{I89, IKR, I91}, 
Witten's Topological Quantum Field Theory \c{Witten88, Witten89, Nash}, 
Gibbons--Hawking's topology change in GR \cite{GH92}, 
and the Author's `topologenesis' continuation of Isham's work in this direction \c{ASoS, ABook, Forthcoming}.
Background Independence moreover bears close relation to Quantum Gravity and in particular Problem of Time themes, as characterized by Wheeler, DeWitt, Kucha\v{r} and Isham, 
and subsequently reviewed by the Author \cite{Battelle, DeWitt67, Kuchar92, I93, APoT1, APoT2, FileR, APoT3} 
(noting \c{ABook} as the most recent and by far most extensive work on this subject). 
Topological Background Independence goes beyond the more usual \cite{Kuchar92, I93, KieferBook, RovelliBook} metric-or-differential-geometry level 
of Background Independence of Geometrodynamics and Loop Quantum Gravity, albeit for now just with various kinds of simple models.  
The current article has added a useful further class of model to these considerations.  

\m 

\n{\bf Acknowledgments} I thank Chris Isham and Don Page for discussions about configuration space topology, geometry, quantization and background independence. 
I also thank Jeremy Butterfield and Christopher Small for encouragement. 
I thank Don, Jeremy, Enrique Alvarez, Reza Tavakol and Malcolm MacCallum for support with my career. 
Finally, writing up this work was possibilitated by a pro-offerer of pineapples' patience and friendship.  
Thank you.  


\end{document}